\documentclass[3p,english,12pt,times,sort&compress,nopreprintline]{elsarticle}
\usepackage{fullpage}

\usepackage[utf8]{inputenc}
\usepackage[T1]{fontenc}
\usepackage{babel}
\usepackage{array}
\usepackage{multirow}
\usepackage{amsmath}
\usepackage{amsthm}
\usepackage{amssymb}
\usepackage{nicefrac}
\usepackage{graphicx}
\usepackage[unicode=true]{hyperref}
\usepackage{xcolor}
\usepackage{textgreek}
\usepackage{comment}
\usepackage[title]{appendix}
\usepackage{subfig}
\usepackage{longtable}
\usepackage{multirow}
\usepackage{booktabs}
\usepackage{caption}
\usepackage{array}
\usepackage{calc}
\usepackage{etoolbox}

\hypersetup{
  colorlinks=true,
  linkcolor=blue,
  filecolor=magenta,
  urlcolor=cyan,
}

\makeatletter
  \newcommand{\LoadPackageOnce}[2][]{%
    \@ifpackageloaded{#2}{%
        \@latex@warning{:: Package #2 already loaded}%
    }{%
        \@latex@warning{:: Package #2 autoloaded with args #1}%
        \usepackage[#1]{#2}}}
\makeatother
\LoadPackageOnce{amsmath}
\LoadPackageOnce{amsthm}
\LoadPackageOnce{amssymb}
\LoadPackageOnce{enumitem}
\LoadPackageOnce{babel}
\LoadPackageOnce{xparse}

\makeatletter
  \NewDocumentCommand{\DefineTheorem}{ m O{} m }{%
    \ifcsmacro{#1}{
      \@latex@warning{:: Theorem #1 is being replaced}%
      \renewtheorem{#1}[#2]{#3}%
    }{%
      \@latex@warning{:: Theorem #1 is being created}%
      \newtheorem{#1}[#2]{#3}%
    }%
  }
  \DefineTheorem{theorem}{\protect\theoremname} %% Alias for thm
  \theoremstyle{plain}
  \DefineTheorem{thm}{\protect\theoremname}
  \def\theoremname{Theorem}

  \theoremstyle{plain}
  \DefineTheorem{cor}[thm]{\protect\corollaryname}
  \def\corollaryname{Corollary}

  \theoremstyle{plain}
  \DefineTheorem{lem}[thm]{\protect\lemmaname}
  \def\lemmaname{Lemma}

  \theoremstyle{plain}
  \DefineTheorem{conjecture}[thm]{\protect\conjecturename}
  \def\conjecturename{Conjecture}

  \theoremstyle{definition}
  \DefineTheorem{defn}[thm]{\protect\definitionname}
  \def\definitionname{Definition}

  \theoremstyle{definition}
  \DefineTheorem{example}[thm]{\protect\examplename}
  \def\examplename{Example}

  \theoremstyle{definition}
  \DefineTheorem{problem}[thm]{\protect\problemname}
  \def\problemname{Problem}

  \theoremstyle{definition}
  \DefineTheorem{xca}[thm]{\protect\exercisename}

  \theoremstyle{definition}
  \DefineTheorem{sol}[thm]{\protect\solutionname}
  \def\solutionname{Solution}

  \theoremstyle{remark}
  \DefineTheorem{rem}[thm]{\protect\remarkname}
  \def\remarkname{Remark}

  \theoremstyle{remark}
  \DefineTheorem{claim}[thm]{\protect\claimname}
  \def\claimname{Claim}

  \newlist{casenv}{enumerate}{4}
  \setlist[casenv]{leftmargin=*,align=left,widest={iiii}}
  \setlist[casenv,1]{label={{\itshape\ \casename \arabic*.}},ref=\arabic*}
  \setlist[casenv,2]{label={{\itshape\ \casename \roman*.}},ref=\roman*}
  \setlist[casenv,3]{label={{\itshape\ \casename\ \alph*.}},ref=\alph*}
  \setlist[casenv,4]{label={{\itshape\ \casename \arabic*.}},ref=\arabic*}
  \def\casename{Case}

  \theoremstyle{plain}
  \DefineTheorem{fact}[thm]{\protect\factname}
  \def\factname{Fact}

\makeatother 
\usepackage{aliascnt}
\makeatletter
\let\lem\@undefined
\let\endlem\@undefined
\makeatother
\theoremstyle{thm}
\newaliascnt{lem}{thm}
\newaliascnt{onlylem}{thm}
\newtheorem{lem}[onlylem]{Lemma}
\aliascntresetthe{onlylem}
\aliascntresetthe{lem}

\makeatletter
  \def\maxwidth{\ifdim\Gin@nat@width>\linewidth\linewidth\else\Gin@nat@width\fi}
  \def\maxheight{\ifdim\Gin@nat@height>\textheight\textheight\else\Gin@nat@height\fi}
  \setkeys{Gin}{width=\maxwidth,height=\maxheight,keepaspectratio}
\makeatother
\makeatletter
  \newcommand{\orcidID}[1]{\href{https://orcid.org/#1}{\protect\includegraphics[width=10pt,scale=0.02]{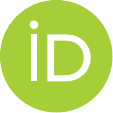}}}
\makeatother
\makeatletter

  \LoadPackageOnce{babel}
  \addto\extrasenglish{

  }
\makeatother
\begin{document}

\begin{frontmatter}

\title{Statistical Inference for Modulation Index in Phase-Amplitude Coupling}

\author[uci]{Marco Antonio Pinto Orellana \orcidID{0000-0001-6495-1305}}
\author[kaust]{Hernando Ombao \orcidID{0000-0001-7020-8091}}
\author[uci]{Beth Lopour \orcidID{0000-0003-4233-4802} \corref{cor}}

\cortext[cor]{Corresponding author}

\address[uci]{Biomedical Engineering Department. University of California, Irvine}
\address[kaust]{Statistics Program. King Abdullah University of Sciences and Technology (KAUST)}

\begin{abstract}
Phase-amplitude coupling is a phenomenon observed in several neurological processes, where the phase of one signal modulates the amplitude of another signal with a distinct frequency.
The modulation index (MI) is a common technique used to quantify this interaction by assessing the Kullback-Leibler divergence between a uniform distribution and the empirical conditional distribution of amplitudes with respect to the phases of the observed signals. 
The uniform distribution is an ideal representation that is expected to appear under the absence of coupling. 
However, it does not reflect the statistical properties of coupling values caused by random chance.
In this paper, we propose a statistical framework for evaluating the significance of an observed MI value based on a null hypothesis that a MI value can be entirely explained by chance.
Significance is obtained by comparing the value with a reference distribution derived under the null hypothesis of independence (i.e., no coupling) between signals.
We derived a closed-form distribution of this null model, resulting in a scaled beta distribution.
To validate the efficacy of our proposed framework, we conducted comprehensive Monte Carlo simulations, assessing the significance of MI values under various experimental scenarios, including amplitude modulation, trains of spikes, and sequences of high-frequency oscillations. Furthermore, we corroborated the reliability of our model by comparing its statistical significance thresholds with reported values from other research studies conducted under different experimental settings.
Our method offers several advantages such as meta-analysis reliability, simplicity and computational efficiency, as it provides p-values and significance levels without resorting to generating surrogate data through sampling procedures.

\end{abstract}


\end{frontmatter}

\section{Introduction}

Phase-amplitude coupling (PAC) is a wide-ranging phenomenon in the brain in which the amplitude of a high-frequency oscillation interacts with the phase of a low-frequency oscillation \citep{PhaseAmplitudeCouplingGeneral-Bergmann-2018, QuantificationPhaseAmplitudeCoupling-Hulsemann-2019, PhaseAmplitudeCoupling-Tort-2010}.It is thought to facilitate communication processes that integrate neural structures or networks \citep{DirectModulationIndex-Scherer-2023, ProperEstPhase-Dvorak-2014, TemporalspatialPhaseamplitude-Zhang-2017, FRoleCrossfrequency-Canolty-2010}. This coupling effect has been observed across brain regions, experimental conditions, and species \citep{TransientPhaseamplitude-Martinez-Cancino-2019}. Due to its broad relevance,  a set of different metrics have been developed to quantify this type of cross-frequency interaction \citep{QuantificationPhaseAmplitudeCoupling-Hulsemann-2019}.

PAC methods typically use the analytic signal representation of a time series, which involves decomposing a signal into its instantaneous phase and amplitude components using a Hilbert transform. Then a metric is commonly defined by quantifying some aspect of the relationship between the instantaneous amplitude of a high-frequency signal ($A_H(t)$) and the instantaneous phase of a low-frequency signal
%\footnote{In the notation for the "instantaneous" amp and phase, I think it would be better if we include time "t".}($\phi_L$)
, or its associated phasor $e^{j\phi_L}(t)$. Three metrics are commonly used: phase locking value, mean vector length, generalized linear models, and modulation index.

Phase locking value (PLV) measures the degree of coupling in phase between two signals by calculating the average phasor difference of the low-frequency and high-frequency phases: $\frac{1}{N}\left|\sum_{n=1}^{N}e^{j\left(\phi_H\left(t\right)-\phi_L\left(t\right)\right)}\right|$ \citep{PhaseSynchrony-Lachaux-1999,InfraslowOscModulate-Vanhatalo-2004,PhaseAmplitudeReset-Mormann-2005}. Even though that PLV only evaluates phase-phase interactions, it is often assumed to be part of the PAC methods \cite{DirectModulationIndex-Scherer-2023}.

Another common PAC metric includes Mean vector length (MLV). This method estimates the average magnitude of the phasor with amplitude $A_H$ and the phase $\phi_L$: $\left|\frac{1}{N}\sum_{n=1}^{N}A_{H}e^{j\phi_{L}\left(t\right)}\right|$ to measure their coupling as result of mutual variability \citep{HighGammaPower-Canolty-2006}. Under no coupling (null hypothesis), the MLV score should be negligible, although nonzero.

Alternatively, we can formulate flexible generalized linear models (GLMs) that represents the interactions between instantaneous phases and amplitudes. Typically, under GLM, the log of the mean amplitude $A_H$ is expressed as a linear combination of spline functions of the phase $\phi_L$ and possibly an intercept. The coefficients of this model are used to quantify the cross-frequency coupling as the scaled maximum difference between the predicted amplitudes and the observed mean of $A_H$ \citep{AssessmentCrossfrequencyCoupling-Kramer-2013}.

Modulation index (MI) is an alternative quantity that measures how different is the probability of observing an amplitude $A_H$ under a phase $\phi_L$ from a uniform distribution. A fundamental question is whether an observed MI value is "statistically significant" to support the presence of PAC in the signals. Due to the complexity of calculating these values, there are limited theoretical statistical frameworks for conducting this assessment. This paper aims to fill that gap by developing a  rigorous framework.
%\footnote{Marco: can we rewrite as follows: Modulation index (MI) is a quantity that measures ... The key question that arises is whether or not an observed MI-value is "statistically significant" to conclude the presence or PAC. There is currently limited framework ... Thus, the contribution of this paper is .... Then go into the details below on constructing the histogram (empirical distribution) etc ...}
MI requires constructing a histogram associating the average $\phi_L$ observed due to $A_H$. Then, MI is calculated as the Kullback-Leiber divergence between the uniform distribution and the normalized histogram \citep{DynamicCrossfrequencyCouplings-Tort-2008, PhaseAmplitudeCoupling-Tort-2010}. In the absence of coupling, the chances of seeing amplitudes at any phase are equal, resulting in a null MI value. Compared with other approaches, MI has a lower false positive rate than PLV or other coupling methods \citep{AddressingPitfallsPhaseAmplitude-Jurkiewicz-2021}. Moreover, it is a potential biomarker of seizure onset zones, as it was shown by Zhang et al.~and Guirguis et al.~in electrocorticogram data from patients with temporal lobe epilepsy \citep{TemporalspatialPhaseamplitude-Zhang-2017, RoleDeltamodulatedHigh-Guirgis-2013}.

Despite the relevance of the MI in quantifying phase-amplitude coupling, no statistical framework is available to assess the significance of the MI scores, as compared to other techniques such as GLM \citep{ProperEstPhase-Dvorak-2014,StatisticalFrameworkAssess-Nadalin-2019}. Gothel et al.~have highlighted the need to rely on surrogate data to obtain estimations of uncertainty due to that reason \citep{AddressingPitfallsPhaseAmplitude-Jurkiewicz-2021, EvaluationPhaseAmplitudeCoupling-Gohel-2016}. Even though this approach is flexible, it incurs high computational time constraints due to the need to generate simulated signals and estimate their modulation index. Moreover, there has yet to be a consensus on a common method of obtaining surrogate signals. Thus, several types of sampling for generating surrogate data have been proposed, including permuting the individual time points of the original data \citep{PhaseAmplitudeCoupling-Tort-2010}, permuting blocks of the original data (block-resampling) \citep{TimeresolvedPhaseamplitudeCoupling-Samiee-2017, UntanglingCrossfrequencyCoupling-Aru-2015}, permuting solely the phase signal while preserving the amplitude signal \citep{AutomatedLateralizationTemporal-Gautham-2022, EmpiricalPhaseamplitude-Caiola-2019}, or permuting only the amplitude signal \citep{HumanThalamusRegulates-Malekmohammadi-2015}. An alternative proposed method is to create surrogates from Gaussian white noise that resembles the same length and variance as the original data \citep{EvaluationPhaseAmplitudeCoupling-Gohel-2016}. Moreover, the wide variation in these approaches to create a null distribution does not allow direct comparisons between the significance across different studies.

This paper introduces a statistical method for analyzing the modulation index in the context of phase-amplitude coupling. Specifically, we propose a closed-form approximation that allows a complete statistical characterization of the distribution of MI values caused by white noise. This approach eliminates the need for surrogate sampling procedures and enhances computational efficiency. Moreover, this model is used to develop a hypothesis test to evaluate the significance of modulation indices. We explore the potential use of our method by testing its ability to provide a threshold for coupling created in amplitude modulation scenarios and trains of spikes or high-frequency oscillations. In addition, we provide a reference source code implementation in R, Python, and Matlab.

\section{Statistical properties of the modulation index}

\hypertarget{sec:modulation-index}{%
\subsection{Modulation index}\label{sec:modulation-index}}

\begin{figure}
\hypertarget{fig:workflow}{%
\centering
\includegraphics{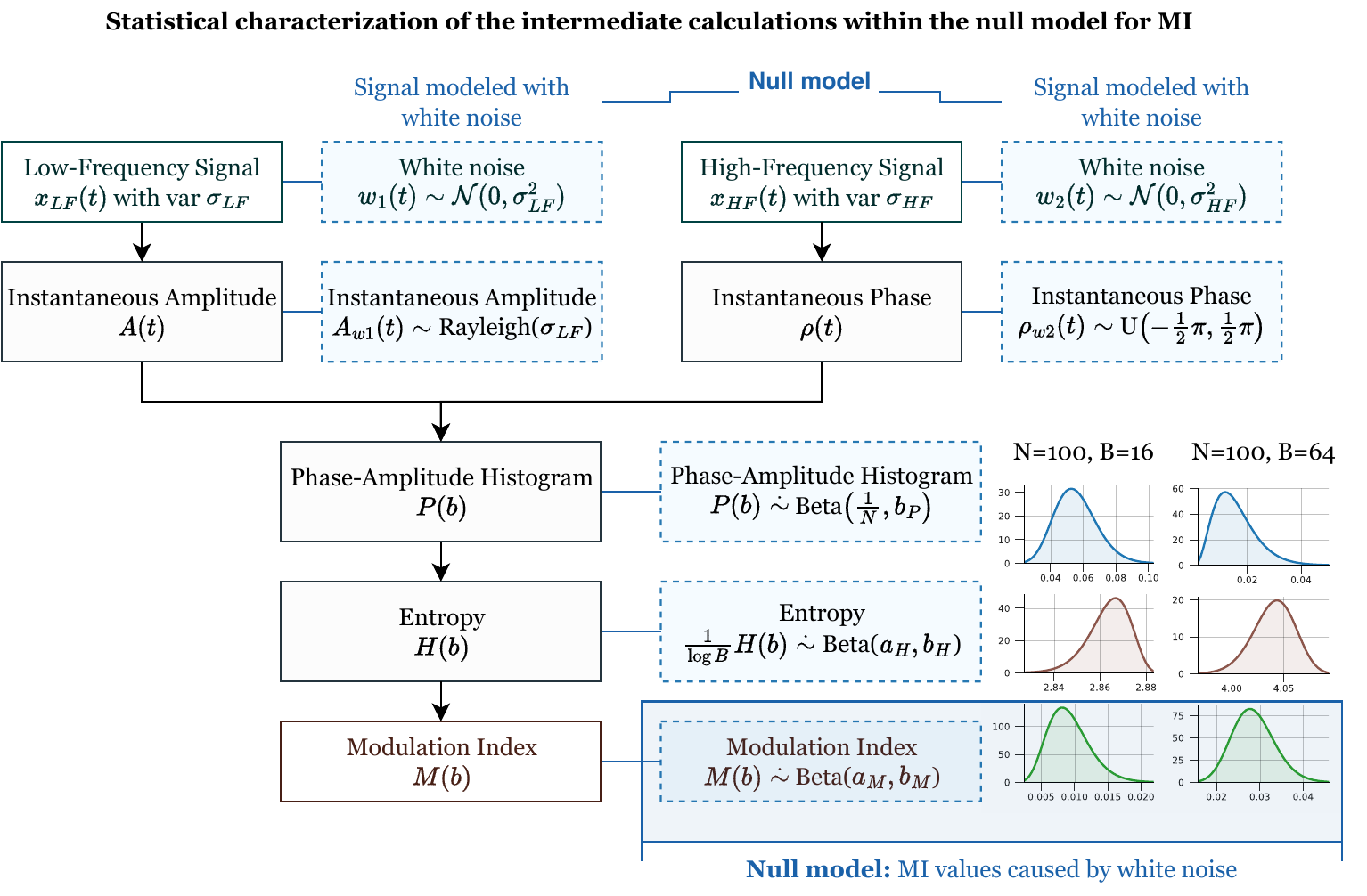}
\caption{Overview of the proposed null model through the intermediate steps to calculate the modulation index.}\label{fig:workflow}
}
\end{figure}

In this subsection, we outline the method to calculate MI proposed by Tort et al.~in \citep{PhaseAmplitudeCoupling-Tort-2010}. This procedure will serve as a basis for developing the statistical behavior of the metric when pure random components are introduced as input while retaining the same notation (as it is depicted in \autoref{fig:workflow}).

To estimate MI, we begin with two signals, $x_{LF}(n)$ and $x_{HF}(n)$, with their power spectra concentrated around low- and high-frequency components, respectively. Both signals can be extracted from the same EEG channel, or they can be extracted from different channels. The analytic signal of each signal is then determined by adding its Hilbert transform (HT) rotated $\frac{\pi}{2}$ in the complex plane. Note that the HT of a signal $x(n)$ is the convolution between the signal itself and a Dirichlet kernel $h(n)$: $\hat{x}(n)=x(n)\ast h(n)$, where the kernel $h(n)$ returns zero when $n$ is even and $\frac{2}{\pi n}$.

Next, we obtain the instantaneous phase of the low-frequency signal (the angle of its analytic signal) and the instantaneous amplitude of the high-frequency signal (the magnitude of its analytic signal). From this, we construct a bivariate signal ${(\phi_{LF}(n),A_{HF}(n))}$ which contains the cross-frequency coupling interactions across time. We then divide the phase space between $0$ and $2\pi$ into $B$ bins and calculate the mean value of the amplitude conditional to the phases being restricted in the interval $[b\frac{2\pi}{B},(b+1)\frac{2\pi}{B})$ for each bin $b$. We normalize these expected values to construct a phase-amplitude histogram $P(b)$ as follows: %
\begin{equation}%
{P(b)=\frac{\mathbb{E}\left[A_{LF}(n)\vert\phi_{HF}\in\left(2\frac{\pi}{B},(b+1)\frac{2\pi}{B}\right)\right]}{\sum_{b=0}^{B-1}\mathbb{E}\left[A_{LF}(n)\vert\phi_{HF}\in\left(2\frac{\pi}{B},(b+1)\frac{2\pi}{B}\right)\right]}\label{eq:pseudo-prob}}%
\end{equation}%

Afterward, we calculate the entropy of this ``pseudo-distribution'' as a measure of the average information contained by this distribution, i.e., the probability of observing a value multiplied by the logarithm base-$K$ of its probability: %
\begin{equation}%
{H_P=-\sum_{b=0}^{B-1}P(b)\log_{K}P(b)\label{eq:entropy}}%
\end{equation}%

It is important to note that discrete Shannon entropy $H$ is bounded between $0$ and $\log_{K}(B)$. When all recorded instantaneous phases lie in a single ``bin'', entropy will be zero, indicating that there is no randomness in the distribution. Conversely, entropy will reach its maximum value when the pseudo-probability behaves as a uniform distribution where all phases have a similar probability of observing the same instantaneous amplitudes. The entropy of such a uniform distribution is $H_{U}=\log_{K}B$, which is expected to happen when there is no coupling between the low and high frequencies.

Finally, the modulation index is defined as the Kullback-Leibler divergence (KLD) between the uniform distribution $U$ and the pseudo-probability $P$ normalized by $\log_{K}B$: %
\begin{equation}%
{\rho_{\text{MI}}=\frac{1}{\log_{K}B}D_{KL}(P\vert U)=\frac{1}{\log_{K}B}(H_{U}-H_{P})=1-\frac{1}{\log_{K}B}H_{P}\label{eq:mod-index}}%
\end{equation}%

While KLD is not generally bounded to a specific range for statistical distributions, the MI formulation restricts its range to be $(0,1)$.  However, it is implausible for MI to reach its maximum, as this value may not indicate conclusive coupling but likely a lack of randomization in the data, as discussed previously. This constrained interval, but counter-intuitive interpretation, has led to discussions on the accuracy of categorizing MI as bounded \citep{DirectModulationIndex-Scherer-2023}.

\subsection{Analytic signal of white noise}

The Hilbert transform of a finite signal $x\left(n\right)$ can be formulated as a causal filter, i.e., as a weighted sum of the past values of the signal ( \autoref{sec:analytic-signal-as-sum}): %
\begin{equation}%
{\hat{x}(n)=\sum_{\ell=1}^{N-n}x\left(n+\ell\right)h\left(\ell\right)-\sum_{\ell=1}^{n}x\left(n-\ell\right)h\left(\ell\right)}%
\end{equation}%
 where $N$ is the size of the sample.

Under the assumption that $x(n)$ is a realization of white noise, $x(n)\sim\mathcal{N}\left(0,\sigma_{x}^{2}\right)$, it is also uncorrelated with previous values $x(n)\perp x\left(m\right)\,\forall n\ne m$. Due to the properties of Gaussian random variables, we can denote that $\hat{x}(n)$ also follows a normal distribution with zero mean and variance $\sigma_{H}^{2}=\kappa(n)\sigma_{x}^{2}$ where the scaling factor $\kappa$ is %
\begin{equation}%
{\kappa(n)=\sum_{\ell=1}^{N-n}h^{2}\left(\ell\right)-\sum_{\ell=1}^{n}h^{2}\left(\ell\right)}%
\end{equation}%

For sample sizes $N>100$, the factor $\kappa(n)$ is approximately one (\autoref{sec:sum-of-weights-HT}). Then, the Hilbert transform of a realization of white noise will be a Gaussian distribution with approximately the same variance, %
\begin{equation}%
{\hat{x}(n)\overset{\text{approx}}{\sim}\mathcal{N}\left(0,\sigma_{x}^{2}\right)}%
\end{equation}%
 .

\begin{lem}[Independence of WN and its Hilbert transform] \label{thm:uncorrelatedness} The Hilbert transform of a sampled white noise realization is statistically independent of the realization.

\end{lem}

\begin{proof} Let us evaluate the correlation between $x$ and $\hat{x}$ by specifying the Hilbert transform as a linear combination of previous and current values: %
{\begin{align}\text{cov}\left(x(n),\hat{x}(n)\right) & =\text{cov}\left(x(n),\sum_{\ell=1}^{N-n}x\left(n+\ell\right)h\left(\ell\right)-\sum_{\ell=1}^{n}x\left(n-\ell\right)h\left(\ell\right)\right)\\
 & =\sum_{\ell=1}^{N-n}h\left(\ell\right)\text{cov}\left(x(n),x\left(n+\ell\right)\right)-\sum_{\ell=1}^{n}h\left(\ell\right)\text{cov}\left(x(n),x\left(n-\ell\right)\right)
\end{align}}%
~ \end{proof}

Because $x(n)$ is uncorrelated with $x(m)$ for different time points $n$ and $m$, the covariance between white noise and its Hilbert transform $\text{cov}\left(x(n),\hat{x}(n)\right)=0$. Then, given that both are normally distributed, they are also independent.

\begin{lem}[WN instantaneous amplitude] \label{thm:magnitude} The magnitude $A_{x}(n)$ of the analytic signal of a zero-mean white noise with variance $\sigma_{x}^{2}$ can be approximated with a Rayleigh random variable, %
\begin{equation}%
{A_{x}(n)\overset{approx}{\sim}\text{Rayleigh}\left(\sigma_{x}\right)}%
\end{equation}%
 with mean $\mathbb{E}\left[A_{x}(n)\right]\approx\sigma_{x}\sqrt{\frac{\pi}{2}}$ and variance $\text{var}\left[A_{x}(n)\right]\approx\sigma_{x}^{2}\left(2-\frac{\pi}{2}\right)$

\end{lem}

\begin{proof} The Hilbert transform of a Gaussian variable is approximately an independent Gaussian random variable (\autoref{thm:uncorrelatedness}). The sum of the squared values of the signal and its Hilbert transform, normalized by their variances, follows a chi-squared distribution, %
\begin{equation}%
{\frac{1}{\sigma_{x}^{2}}x^{2}(n)+\frac{1}{\sigma_{x}^{2}}\hat{x}^{2}(n)\sim\chi_{2}^{2}}%
\end{equation}%
 Given that the variances are equal, the absolute value of the analytic signal follows a Rayleigh distribution (or a weighted chi-distribution with two degrees of freedom), \begin{equation}{A_{x}(n)=\sqrt{x^{2}(n)+\hat{x}^{2}(n)}\overset{approx}{\sim}\sigma_{x}\chi_{2}=\text{Rayleigh}\left(\sigma_{x}\right)}\end{equation} As a consequence, the mean and variance of the amplitude of the analytic signal is described by 
 {\begin{align}\mathbb{E}\left[A_{x}(n)\right] & \approx\sigma_{x}\sqrt{\frac{\pi}{2}}\\
\text{var}\left[A_{x}(n)\right] & \approx\sigma_{x}^{2}\left(2-\frac{\pi}{2}\right).
\end{align}}%
~ \end{proof}

\begin{lem}[WN instantaneous phase] \label{thm:phase} The phase of a white noise signal follows a uniform distribution in $[-\frac{\pi}{2},\frac{\pi}{2}]$, and it is, therefore, independent of the variance of the signal.

\end{lem}

\begin{proof} Recall that white noise and its Hilbert transform are independent and normally distributed with the same variance. Then, the ratio between them will be a Cauchy distribution: %
\begin{equation}%
{\frac{\hat{x}(n)}{x(n)}\sim\text{Cauchy}(0,1)}%
\end{equation}%
 By definition, a Cauchy distribution does not have a finite mean or variance. However, it has a relevant property that explains that its arc-tangent value has a uniform distribution \citep{RatiosNormalVariables-Marsaglia-1965}: \begin{equation}{\phi(n)=\arctan\left(\frac{\hat{x}(n)}{x(n)}\right)\sim\text{U}\left(-\frac{\pi}{2},+\frac{\pi}{2}\right)}\end{equation} Therefore, the instantaneous phase for white noise is uniformly distributed.~

\end{proof}

\subsection{Proposed null model}

Let us consider the modulation index obtained by analyzing the cross-frequency coupling between two white noise signals. The reasoning behind this is that white signals have a spectrum power that is uniformly distributed across all frequencies, meaning they have the same chance of having high- and low-frequency components in the target frequency ranges of the experiment.

Let $w_{1}(n)$ and $w_{2}(n)$ be samples from independent white noise with variances $\sigma_{1}^{2}$ and $\sigma_{2}^{2}$, respectively. Taking into account the independence of the compared signals, the conditional distribution of amplitude with respect to the phases will be a Rayleigh distribution that is independent of the phase itself ($A_{w_{1}}\perp\phi_{w_{2}}$): %
\begin{equation}%
{A_{w_{1}}(n)\vert\phi_{w_{2}}(n)=A_{w_{1}}(n)\overset{approx}{\sim}\text{Rayleigh}\left(\sigma_{1}\right)\label{eq:conditional-independence}}%
\end{equation}%

Before exploring the white noise behavior in the phase-amplitude histogram, let us analyze the average of $M$ independent random variables that follows a Rayleigh distribution.

\begin{lem}[Average of random Rayleigh samples] \label{thm:average-Rayleigh-samples} Let $A_{1},A_{2},\ldots,A_{M}$ be independent random variables with a Rayleigh distribution, where $M$ is also a random variable that follows a binomial distribution. Let $p$ be the probability of success, $M_{0}$ be the number of trials for the binomial distribution, and let $s$ be the scale parameter of the Rayleigh distribution. The mean and variance of the sum $S=\frac{1}{M}\left(A_{1}+A_{2}+\ldots+A_{M}\right)$ are given by %
{\begin{align}
\mathbb{E}\left[S\right] & =s\sqrt{\frac{\pi}{2}}\\
\text{var}\left(S\right) & \approx s^{2}\frac{\pi}{2}\left(1+\frac{\frac{4}{\pi}-1}{M_{0}p}\left(1+\frac{1-p}{M_{0}p}\left(1-\frac{1-2p}{2M_{0}p}\right)\right)\right)\end{align}}%
 \end{lem}

\begin{proof} To derive the expectation of the sum by using the law of total expectation, %
\begin{equation}%
{\mathbb{E}\left[S\right]=\mathbb{E}_{M}\left[\mathbb{E}\left[\frac{1}{M}\sum_{m=1}^{M}A_{m}\vert M\right]\right]=\mathbb{E}_{M}\left[s\sqrt{\frac{\pi}{2}}\right]=s\sqrt{\frac{\pi}{2}}}%
\end{equation}%

Identically, we can obtain variance by the law of total variance, %
{\begin{align}
\text{var}\left(S\right) & =\mathbb{E}_{M}\left[\text{var}\left(\frac{1}{M}\sum_{m=1}^{M}A_{m}\vert M\right)\right]+\text{var}_{M}\left(\mathbb{E}\left[\frac{1}{M}\sum_{m=1}^{M}A_{m}\vert M\right]\right)\\
 & =\mathbb{E}_{M}\left[\frac{1}{M}s^{2}\frac{4-\pi}{2}\right]+\text{var}_{M}\left(s\sqrt{\frac{\pi}{2}}\right)\\
 & =s^{2}\,\frac{4-\pi}{2}\mathbb{E}_{M}\left[\frac{1}{M}\right]\end{align}}%
  By Taylor expansion of $1/M$, we can approximate the expectation of the inverse of $M$ based on its mean $\mu_M$ and variance $\sigma_M^2$, %
 {\begin{align}
\mathbb{E}\left[\frac{1}{M}\right] & \approx\mu_{M}^{-1}+\sigma_{M}^{2}\mu_{X}^{-3}-\frac{1}{2}\mathbb{E}\left[\left(m-\mu_{M}\right)^{3}\right]\mu_{X}^{-4}\end{align}}%
 Recall that the third central moment of $M$ is $\mathbb{E}[(m-\mu_{M})^{3}]=M_{0}p(1-p)(1-2p)=\sigma_{X}^{2}(1-2p)$ %
{\begin{align}
\mathbb{E}_{M}\left[\frac{1}{M}\right] & \approx\frac{1}{M_{0}p}+M_{0}p\left(1-p\right)\frac{1}{M_{0}^{3}p^{3}}-M_{0}p\left(1-p\right)\frac{1-2p}{2M_{0}^{4}p^{4}}\\
 & =\frac{1}{M_{0}p}\left(1+\frac{1-p}{M_{0}p}\left(1-\frac{1-2p}{2M_{0}p}\right)\right)\end{align}}%
  Then, finally %
 {\begin{align}
\text{var}\left(S\right) & \approx\frac{s^{2}}{2}\left(\pi+\frac{4-\pi}{M_{0}p}\left(1+\frac{1-p}{M_{0}p}\left(1-\frac{1-2p}{M_{0}p}\right)\right)\right)\end{align}}%
~ \end{proof}

Now, let us construct a phase-amplitude histogram with $B$ bins (Equation~\ref{eq:pseudo-prob}). As discussed in \autoref{thm:uncorrelatedness}, phase and amplitude are independent, and each amplitude has an equal probability $\frac{1}{B}$ of being selected for one of the bins as the phase is uniformly distributed. Therefore, the sum of elements in each bin has the same expectation and variance as described in \autoref{thm:average-Rayleigh-samples} with $p=\frac{1}{B}$.

Using the moments of the averages inside each bin, we can construct a phase-amplitude histogram, $P$, as the proportions of an average $S_{i}$ of Rayleigh-distributed random variables, with respect to the total average $T=S_{1}+S_{2}+\ldots+S_{B}$.

\begin{lem}[Phase-amplitude histogram moments] \label{thm:pseudo-prob-moments} Let $A_{1},A_{2},\ldots,A_{N}$ be $N$ Rayleigh-distributed random variables, and assume they are uniformly distributed among $B$ groups. Then, let $S_{b}$ be the average of all variables in group \textbf{$b$} and $T$ the overall average $T=S_{1}+S_{2}+\ldots+S_{B}$. Then, the proportion $p_{b}=\frac{S_{b}}{T}$ has a mean and variance approximately described by %
{\begin{align}
\mu_p = \mathbb{E}\left(p_{b}\right) & \approx\frac{1}{B}\\
\sigma_p^2 = \text{var}\left(p_{b}\right) & \approx B\left(1-\frac{1}{B}\right)\left(\frac{1}{B^{3}}+\frac{\frac{4}{\pi}-1}{N}\left(\frac{1}{B^{2}}+\frac{1-\frac{1}{B}}{N}\left(\frac{1}{B}-\frac{1-\frac{2}{B}}{2N}\right)\right)\right)\end{align}}%
 \end{lem}

\begin{proof} Let us study an approximation to the moments of the proportion $\frac{S_{b}}{T}$ by using a Taylor expansion around the $\frac{\mu_{S}}{\mu_{T}}$, %
{\begin{align}
\mathbb{E}\left[\frac{S_{b}}{T}\right] & \approx\frac{\mu_{S}}{\mu_{T}}+\sigma_{T}^{2}\frac{\mu_{S}}{\mu_{T}^{3}}-\sigma_{S,T}\frac{1}{\mu_{T}^{4}}\\
\text{var}\left(\frac{S_{b}}{T}\right) & \approx\sigma_{S}^{2}\frac{1}{\mu_{T}^{2}}+\sigma_{S}^{2}\frac{\mu_{S}^{2}}{\mu_{T}^{4}}-2\sigma_{S,T}\frac{\mu_{S}}{\mu_{T}^{3}}\label{eq:taylor-approximation}\end{align}}%
 where $\mu_{S}$ and $\sigma_{S}^{2}$ are the mean and variance of the average $S_{b}$ inside a bin, and $\mu_{T}$ and $\sigma_{T}^{2}$ are the mean and variance of the sum of averages $T$.

The probability of a variable being placed into a group is $\frac{1}{B}$. Then, the size of each group $M_{b}$ can be modeled with a binomial distribution. Inside each group, the first and second central moments of the average $S_{b}=\frac{1}{M_{b}}\sum_{k=1}^{M_{b}}A_{k}$ is described in \autoref{thm:average-Rayleigh-samples}, and it will be identical for all groups.

The expectation of the total average $T$ is the sum of the means of $S_{b}$, $\mu_{T}=\mathbb{E}\left[T\right]=\sum_{b=1}^{B}\mathbb{E}\left[S_{b}\right]=B\mu_{b}$. In addition, given that the averages $S_{k}$ are uncorrelated among themselves, the variance of $T$ can also be described as a function of the variance of $S_{b}$: $\sigma_{T}^{2}=\text{var}\left(T\right)=\sum_{b=1}^{B}\text{var}\left(S_{b}\right)=B\sigma_{b}^{2}$.

Furthermore, the covariance between the numerator and denominator of the proportion will be equal to the variance of the numerator, $\sigma_{S,T}=\text{cov}\left(S_{b},\sum_{k=1}^{B}S_{k}\right)=\text{cov}\left(S_{b},S_{b}\right)+\text{cov}\left(S_{b},\sum_{k=1,k\ne b}^{B}S_{k}\right)=\sigma_{S}^{2}$. Then, replacing $\mu_{S}$, $\mu_{T}$, $\sigma_{S}^{2}$, $\sigma_{T}^{2}$ and $\sigma_{S,T}$ into Equation~\ref{eq:taylor-approximation}, %
\begin{align}
\mathbb{E}\left(\frac{S_{b}}{T}\right) & \approx\frac{1}{B}+B\sigma_{S}^{2}\frac{1}{B^{3}\mu_{S}^{2}}-\sigma_{S}^{2}\frac{1}{B\mu_{S}^{2}}=\frac{1}{B}\nonumber\\
\text{var}\left(\frac{S_{b}}{T}\right) & \approx\sigma_{S}^{2}\frac{1}{B^{2}\mu_{S}^{2}}+B\sigma_{S}^{2}\frac{\mu_{S}^{2}}{B^{4}\mu_{S}^{4}}-2\sigma_{S}^{2}\frac{\mu_{S}}{B^{3}\mu_{S}^{3}}\nonumber\\
 & =\left(\frac{1}{B^{2}}-\frac{1}{B^{3}}\right)\frac{\sigma_{S}^{2}}{\mu_{S}^{2}}=\frac{1}{B^{2}}\left(1-\frac{1}{B}\right)\frac{\sigma_{S}^{2}}{\mu_{S}^{2}}\end{align}%
  Thus, the mean of the proportion is reciprocal to the number of groups $B$, and the variance is proportional to the squared coefficient of variation of $S$. The latter factor, can be found in the moments in \autoref{thm:average-Rayleigh-samples}, %
 \begin{equation}%
 {\frac{\sigma_{S}^{2}}{\mu_{S}^{2}}\approx\frac{\frac{\pi}{2}\left(1+\frac{\frac{4}{\pi}-1}{\frac{N}{B}}\left(1+\frac{1-\frac{1}{B}}{\frac{N}{B}}\left(1-\frac{1-2\frac{1}{B}}{2\frac{N}{B}}\right)\right)\right)}{\frac{\pi}{2}}=B^{3}\left(\frac{1}{B^{3}}+\frac{\frac{4}{\pi}-1}{N}\left(\frac{1}{B^{2}}+\frac{1-\frac{1}{B}}{N}\left(\frac{1}{B}-\frac{1-\frac{2}{B}}{2N}\right)\right)\right)}\end{equation}

Finally, the variance of the proportion $p_{b}$ can be described as %
\begin{equation}%
{\text{var}\left(\frac{S_{b}}{T}\right)\approx B\left(1-\frac{1}{B}\right)\left(\frac{1}{B^{3}}+\frac{\frac{4}{\pi}-1}{N}\left(\frac{1}{B^{2}}+\frac{1-\frac{1}{B}}{N}\left(\frac{1}{B}-\frac{1-\frac{2}{B}}{2N}\right)\right)\right)}%
\end{equation}%
 Note that in large samples, $N\rightarrow\infty$, the variance only depends on the number of groups $B$, $\text{var}\left(p_{b}\right)\approx\frac{1}{B^{2}}\left(1-\frac{1}{B}\right)$. In addition, the larger the number of groups $B$, the lower the dispersion of the proportion.~ \end{proof}

The phase-amplitude histogram value $p_{k}$ has a domain in the range $[0,1]$, and the sum of all ratios will equal one. Based on these conditions, we propose to model these ratios with a Dirichlet distribution. As a consequence, the marginal distribution for a particular bin $k$ will have a beta distribution. Using the mean and variance for ratios discussed in \autoref{thm:pseudo-prob-moments}, the parameters $a$ and $b$ of the marginal beta distributions will be estimated using the method of moments: %
{\begin{align}
a & =\mu_{p}\left(\frac{\mu_{p}\left(1-\mu_{p}\right)}{\sigma_{p}^{2}}-1\right)\nonumber \\
b & =\left(1-\mu_{p}\right)\left(\frac{\mu_{p}\left(1-\mu_{p}\right)}{\sigma_{p}^{2}}-1\right)\label{eq:pseudo-prob-parameter}\end{align}}%

Now, recall that the entropy of the entire phase-amplitude histogram $P$ is given by %
\begin{equation}%
{H=-\sum_{b=1}^{B}p_{b}\log p_{b}=-\sum_{b=1}^{B}h_{b}}%
\end{equation}%
where $h_{b}$ can be called the ``entropy contribution'' of bin $b$.

Given that $p_{b}$ are identically distributed, the first moments of entropy can be expressed as a function of the entropy contributions, %
{\begin{align}
\mu_H=\mathbb{E}\left[H\right] & =B\mathbb{E}\left[h_{b}\right] \nonumber\\
\sigma_H^2=\text{var}\left(H\right) & =B\text{var}\left(h_{b}\right)+B\left(B-1\right)\text{cov}\left(h_{i},h_{j}\right) \nonumber\\
 & =B\left(\mathbb{E}\left[h_{b}^{2}\right]-\mathbb{E}^{2}\left[h_{b}\right]\right)+B\left(B-1\right)\left(\mathbb{E}\left[h_{i}h_{j}\right]-\left(\mathbb{E}\left[h_{b}\right]\right)^{2}\right)\label{eq:entropy-moment}
 \end{align}}%

\begin{lem} \label{thm:partial-entropy-moments} Let $P$ be a marginally beta-distributed random variable with parameters $a$ and $b$. Let $h=g(P)$ be a function of $P$ such that $g(P)=P\log P$. Then, the first non-central moments of $h$ will be determined by %
{\begin{align}
\mathbb{E}\left[h\right]=\mathbb{E}\left[P\log P\right] & =\frac{a}{a+b}\left(\psi\left(a+1\right)-\psi\left(a+b+1\right)\right)\\
\mathbb{E}\left[h^{2}\right]=\mathbb{E}\left[\left(P\log P\right)^{2}\right] & =\frac{a\left(a+1\right)}{\left(a+b\right)\left(a+b+1\right)}\left(\psi^{(1)}\left(a+2\right)-\psi^{(1)}\left(a+b+2\right)+\left(\psi\left(a+2\right)-\psi\left(a+b+2\right)\right)^{2}\right)\end{align}}%
 where $\psi\left(x\right)$ is the digamma function and $\psi^{(1)}\left(x\right)$ is its first derivative with respect to $x$. \end{lem}

\begin{proof} Let us start by considering the mean of $h$, %
{\begin{align}
\mathbb{E}\left[h\right] & =\frac{1}{\beta\left(a,b\right)}\int_{0}^{1}p\ln p\:p^{a-1}\left(1-p\right)^{b-1}\:dp\end{align}}%
 where $\beta (a,b)$ is the beta function.%

To solve this integral, we can differentiate under the integral sign given that $\frac{\partial p^{a}}{\partial a}=p^{a}\ln p$. Then, %
{\begin{align}
\mathbb{E}\left[h\right] & =\frac{1}{B\left(a,b\right)}\int_{0}^{1}\frac{\partial}{\partial a}p^{a}\left(1-p\right)^{b-1}\:dp\end{align}}%%
 Later, we can express the mean of $h$ as a function of the mean of $H$, %
{\begin{align}
\mathbb{E}\left[h\right] & =\frac{1}{\beta\left(a,b\right)}\frac{\partial}{\partial a}\left(\mathbb{E}\left[X\right]\beta\left(a,b\right)\right)\\
 & =\frac{1}{\beta\left(a,b\right)}\frac{\partial}{\partial a}\left(\frac{a\beta\left(a,b\right)}{a+b}\right)\end{align}}%

Note that $\beta\left(a,b\right)=\frac{\Gamma\left(a\right)\Gamma\left(b\right)}{\Gamma\left(a+b\right)}$ and $\frac{a}{\left(a+b\right)}\frac{a\Gamma\left(a\right)\Gamma\left(b\right)}{\left(a+b\right)\Gamma\left(a+b\right)}=\frac{\Gamma\left(a+1\right)\Gamma\left(b\right)}{\Gamma\left(a+b+1\right)}=\beta\left(a+1,b\right)$. Then, %
{\begin{align}
\mathbb{E}\left[h\right] & =\frac{\frac{a}{a+b}}{\beta\left(a+1,b\right)}\frac{\partial}{\partial a}\beta\left(a+1,b\right)\\
 & =\frac{a}{a+b}\left(\psi\left(a+1\right)-\psi\left(a+b+1\right)\right)\end{align}}%

Now, consider the second non-central moment of $h$, %
{\begin{align}
\mathbb{E}\left[h^{2}\right] & =\frac{1}{\beta\left(a,b\right)}\int_{0}^{1}p^{2}\ln^{2}p\:p^{a-1}\left(1-p\right)^{b-1}\:dp\end{align}}%
 Similarly, we can differentiate under the integral sign, given that $\frac{\partial p^{a}}{\partial a}=p^{a}\ln p$ and $\frac{\partial^{2}p^{a}}{\partial a^{2}}=p^{a}\ln^{2}p$, %
{\begin{align}
\mathbb{E}\left[h^{2}\right] & =\frac{1}{\beta\left(a,b\right)}\int_{0}^{1}p^{2}\frac{\partial^{2}p^{a-1}}{\partial a^{2}}\left(1-p\right)^{b-1}\:dp\\
 & =\frac{1}{\beta\left(a,b\right)}\frac{\partial^{2}}{\partial a^{2}}\int_{0}^{1}p^{\left(a+2\right)-1}\left(1-p\right)^{b-1}\:dp\end{align}}%
  Therefore, the second moment of $h$ can be described as a second partial derivative of the second moment of $P$, %
 {\begin{align}
\mathbb{E}\left[h^{2}\right] & =\frac{1}{\beta\left(a,b\right)}\frac{\partial^{2}}{\partial a^{2}}\left(\mathbb{E}\left[P^{2}\right]\beta\left(a,b\right)\right)\\
 & =\frac{1}{\beta\left(a,b\right)}\frac{\partial^{2}}{\partial a^{2}}\left(\frac{a\left(a+1\right)}{\left(a+b\right)\left(a+b+1\right)}\beta\left(a,b\right)\right)\end{align}}%

By solving the derivatives, %
{\begin{align}
\mathbb{E}\left[h^{2}\right] & =\frac{a\left(a+1\right)}{\left(a+b\right)\left(a+b+1\right)}\frac{1}{\beta\left(a+2,b\right)}\frac{\partial}{\partial a}\left(\beta\left(a+2,b\right)\left(\psi\left(a+2\right)-\psi\left(a+b+2\right)\right)\right)\\
 & =\frac{a\left(a+1\right)}{\left(a+b\right)\left(a+b+1\right)}\left(\psi^{(1)}\left(a+2\right)-\psi^{(1)}\left(a+b+2\right)+\frac{\left(\psi\left(a+2\right)-\psi\left(a+b+2\right)\right)}{\beta\left(a+2,b\right)}\frac{\partial}{\partial a}\beta\left(a+2,b\right)\right)\\
 & =\frac{a\left(a+1\right)}{\left(a+b\right)\left(a+b+1\right)}\left(\psi^{(1)}\left(a+2\right)-\psi^{(1)}\left(a+b+2\right)+\left(\psi\left(a+2\right)-\psi\left(a+b+2\right)\right)^{2}\right)\end{align}}%
 ~ \end{proof}

\begin{lem} \label{thm:cross-moment-entropy} Let $P_{1},P_{2},\ldots,P_{N}$ be a random vector with a Dirichlet distribution with parameters $a_{1},a_{2},\ldots,a_{N}$. Let $h_{k}=P_{k}\log P_{k}$ be a function of $P_{k}$. The cross-moment between $h_{i}$ and $h_{j}$ is given by,

{\begin{align}
\mathbb{E}\left[h_{i}h_{j}\right] & =-\kappa\left(-\psi^{(1)}\left(A+2\right)\right)\\
 & \quad-\kappa\left(-\psi\left(A+2\right)+\psi\left(a_{j}+1\right)\right)^2\end{align}}%
 where $\kappa$ is a normalization factor $\frac{\Gamma\left(a_{i}+1\right)\Gamma\left(a_{j}+1\right)\Gamma\left(A\right)}{\left(a_{i}\right)\Gamma\left(a_{j}\right)\Gamma\left(A+2\right)}$ and $A=\sum_{b=1}^{N}a_{b}$.
  \end{lem}

\begin{proof} First, define the marginal probability distribution of $(P_{i},P_{j})$ if $P_{1},P_{2},\ldots,P_{N}$ to follow a Dirichlet distribution. Due to the aggregation property of this probability family, the joint distribution can be defined as another Dirichlet distribution, %
\begin{equation}%
{f_{P_{i}P_{j}}\left(p_{i},p_{j}\right)=\xi p_{i}^{a_{i}-1}p_{j}^{a_{j}-1}\left(1-p_{i}-p_{j}\right)^{A^{*}-1}}%
\end{equation}%
 where $A^{*}=\sum_{b=1}^{N}a_{b}-a_{i}-a_{j}$ and $\xi=\frac{\Gamma\left(a_{i}\right)\Gamma\left(a_{j}\right)\Gamma\left(A^{*}\right)}{\Gamma\left(a_{i}+a_{j}+A^{*}\right)}$.

Then, calculate the cross-moment is calculated to be %
{\begin{align}
\mathbb{E}\left[h_{i}h_{j}\right] & =\int\int\left(p_{i}\log p_{i}\right)\left(p_{j}\log p_{j}\right)f_{p_{i}p_{j}}\left(p_{i},p_{j}\right)\,dp_{i}\,dp_{j}\end{align}}%

Again, by differentiating under the integral sign, given that $\frac{\partial}{\partial a}p^{a}=p^{a}\log p$, %
{\begin{align}
\mathbb{E}\left[h_{i}h_{j}\right] & =\frac{1}{\xi}\int\int p_{i}p_{j}\frac{\partial}{\partial a_{i}}p_{i}^{a_{i}-1}\frac{\partial}{\partial a_{j}}p_{j}^{a_{j}-1}\left(1-p_{i}-p_{j}\right)^{A^{*}-1}\,dp_{i}\,dp_{j}\end{align}}%
Then, the cross-moment is described as a function of the cross-moment between $p_{i}$ and $p_{j}$: %
{\begin{align}
\mathbb{E}\left[h_{i}h_{j}\right] & =\frac{1}{\xi}\frac{\partial}{\partial a_{i}}\frac{\partial}{\partial a_{j}}\left(\mathbb{E}\left[p_{i}p_{j}\right]\xi\right)=-\frac{1}{\xi}\frac{\partial}{\partial a_{i}}\frac{\partial}{\partial a_{j}}\frac{\xi a_{i}a_{j}}{A\left(1+A\right)}\end{align}}%

To simplify the notation, define $\eta$ to be %
{\begin{align}
\eta(a_{i},a_{j})=\frac{a_{i}a_{j}}{A\left(1+A\right)}\xi & =\frac{\Gamma\left(a_{i}+1\right)\Gamma\left(a_{j}+1\right)\Gamma\left(A^{*}\right)}{\Gamma\left(A+2\right)}\end{align}}%
 Then, we can differentiate the expressions, %
{\begin{align}
\mathbb{E}\left[h_{i}h_{j}\right] & =-\frac{1}{\xi}\frac{\partial}{\partial a_{i}}\frac{\partial}{\partial a_{j}}\left(\eta(a_{i},a_{j})\right)\\
 & =-\frac{1}{\xi}\frac{\partial}{\partial a_{i}}\left(\eta(a_{i},a_{j})\left(-\psi\left(A+2\right)+\psi\left(a_{j}+1\right)\right)\right)\\
 & =-\frac{\eta(a_{i},a_{j})}{\xi}\left(-\psi^{(1)}\left(A+2\right)\right)\\
 & \quad-\frac{\eta(a_{i},a_{j})}{\xi}\left(-\psi\left(A+2\right)+\psi\left(a_{j}+1\right)\right)^2\end{align}}%

Finally, by simplifying, the cross-moment can be defined as %
{\begin{align}
\mathbb{E}\left[h_{i}h_{j}\right] & =-\frac{\eta}{\xi}\left(-\psi^{(1)}\left(A+2\right)\right)\\
 & \quad-\frac{\eta}{\xi}\left(-\psi\left(A+2\right)+\psi\left(a_{j}+1\right)\right)^2\end{align}}%
 ~ \end{proof}

As mentioned before, entropy has a domain restricted in the range between zero and $\log B$. We propose to model the normalized entropy $H$ as a beta distribution, $\frac{1}{\log_{K}B}H\sim\text{Beta}\left(a_{H},b_{H}\right)$, in which parameters $a_{H}$ and $b_{H}$ can be estimated through the moments of $h_{b}$: %
{\begin{align}
a_{h} & =\mu_{H}D_{H}\nonumber \\
b_{H} & =\left(1-\mu_{H}\right)D_{H}\label{eq:entropy-parameter}\end{align}}%
 where $D_{H}=\frac{\mu_{H}-\mu_{H}^{2}}{\sigma_{H}^{2}}-1$ is a normalization factor, and $\mu_{H}$ and $\sigma_{H}^{2}$ are the first two moments of the sample entropy (Equation~\ref{eq:entropy-moment}).

\begin{theorem} \label{thm:modulation-index-distribution} Let $X_{1},X_{2},\ldots,X_{N}$ and $Y_{1},Y_{2},\ldots,Y_{N}$ be i.i.d. zero-mean Gaussian-distributed random variables with variance $\sigma_{X}^{2}$ and $\sigma_{Y}^{2}$, and serially uncorrelated (white noise signals). The modulation index between the instantaneous phase of $X$ and the instantaneous amplitude of $Y$, obtained by constructing a phase-amplitude histogram with $B$ bins of equal width, can be approximated with a beta distribution, $\rho_{\text{MI}}\sim\text{Beta}\left(b_{H},a_{H}\right)$, with shape parameters $a_{MI}$ and $b_{MI}$ such that, %
{\begin{align}
b_{MI} & =\mu_{H}D_{H}\\
a_{MI} & =\left(1-\mu_{H}\right)D_{H}\\
D_{H} & =\frac{\mu_{H}-\mu_{H}^{2}}{\sigma_{H}^{2}}-1\\
\mu_{H} & =-M_{1}B\\
\sigma_{H}^{2} & =B\left(M_{2}-M_{1}^{2}\right)+B\left(B-1\right)C\\
M_{1} & =\frac{1}{B}h\left(a_{p},1,1\right)\\
M_{2} & =\frac{1}{B}\,\frac{a_{p}+1}{Ba_{p}+1}\left(h^{(1)}\left(a_{p},2,2\right)+h^{2}\left(a_{p},2,2\right)\right)\\
C & =\frac{1}{B}\left(\psi^{(1)}\left(Ba_{p}+2\right)+h^{2}\left(a_{p},1,2\right)\right)\\
a_{p} & =\mu_{p}\left(\frac{\mu_{p}-\mu_{p}^{2}}{\sigma_{p}^{2}}-1\right)\\
\mu_{p} & =\frac{1}{B}\\
\sigma_{p}^{2} & =B\left(1-\frac{1}{B}\right)\left(\frac{1}{B^{3}}+\frac{\frac{4}{\pi}-1}{N}\left(\frac{1}{B^{2}}+\frac{1-\frac{1}{B}}{N}\left(\frac{1}{B}-\frac{1-\frac{2}{B}}{2N}\right)\right)\right)\end{align}}%
 where $h_{k}\left(a,i,j\right)=\psi^{(k)}\left(a+i\right)-\psi\left(Ba+j\right)$and $\psi^{(k)}$ is the k-th derivative of the digamma function. \end{theorem}

\begin{proof} From Equation~\ref{eq:conditional-independence}, the instantaneous phase and amplitude are conditionally independent. Therefore, as previously explored, the entropy of the phase-amplitude histogram $P$, constructed by the procedure described in \autoref{sec:modulation-index}, has the distribution $\frac{1}{\log_{K}B}H\sim\text{Beta}\left(a_{H},b_{H}\right)$ with the parameters described previously. Recall that modulation index $\rho_{\text{MI}}$ (Equation~\ref{eq:mod-index}) %
\begin{equation}%
{\rho_{\text{MI}}=1-\frac{1}{\log_{K}B}H_{P}}%
\end{equation}%
 Due to the reflection symmetry property of the beta probability distribution function, the modulation index has the same probability parameters as entropy but reflected:
 \begin{equation}{\rho_{\text{MI}}\sim\text{Beta}\left(a_{MI},b_{MI}\right)=\text{Beta}\left(b_{H},a_{H}\right)}\end{equation}

From Equation~\ref{eq:entropy-parameter}, those parameters can be expressed as a function of the entropy mean and variance, %
{\begin{align}
b_{MI} & =\mu_{H}D_{H}\\
a_{MI} & =\left(1-\mu_{H}\right)D_{H}\end{align}}%
 where $D_{H}=\frac{\mu_{H}-\mu_{H}^{2}}{\sigma_{H}^{2}}-1$ is a normalization factor.

Those entropy's central moments are described in Equation~\ref{eq:entropy-moment}, %
{\begin{align}
\mu_{H} & =-M_{1}B\\
\sigma_{H}^{2} & =B\left(M_{2}-M_{1}^{2}\right)+B\left(B-1\right)C\end{align}}%
 where $M_{1}$ and $M_{2}$ are the mean and variance of single entropy contributions of the phase-amplitude histogram value $p_{b}$, and $C$ is the cross-moment between entropy at bins $p_{i}$ and $p_{j}$. 
The parameter $a_{p}$ is the shape of the (marginal) beta distribution that modeled $p_{b}$ (Equation~\ref{eq:pseudo-prob-parameter}), %
\begin{equation}%
{a_{p}=\mu_{p}\left(\frac{\mu_{p}-\mu_{p}^{2}}{\sigma_{p}^{2}}-1\right)}\end{equation} where $\mu_{p}$ and $\sigma_{p}^{2}$ are approximations to the mean and variance of the phase-amplitude histogram. 

Recall that $p_1, p_2, \ldots, p_B$ are jointly described by a Dirichlet distribution, then the other shape parameter $b_p$ of the phase-amplitude histogram is given by $b_p=(Ba_p - a_p)$. 
Both shape parameters can be used to define $M_1$ and $M_2$, which have closed-form expressions described in \autoref{thm:partial-entropy-moments} and \autoref{thm:cross-moment-entropy}, %
{\begin{align}
M_{1} & =\frac{a_{p}}{a_{p}+b_{p}}h\left(a_{p},1,1\right)\ =\frac{1}{B}h\left(a_{p},1,1\right)\\
M_{2} & =\frac{a_{p}}{a_{p}+b_{p}}\,\frac{a_{p}+1}{Ba_{p}+1}\left(h^{(1)}\left(a_{p},2,2\right)+h^{2}\left(a_{p},2,2\right)\right)\nonumber\\
&=\frac{1}{B}\,\frac{a_{p}+1}{Ba_{p}+1}\left(h^{(1)}\left(a_{p},2,2\right)+h^{2}\left(a_{p},2,2\right)\right)\\
C &=\frac{a_{p}}{a_{p}+b_{p}}\left(\psi^{(1)}\left(Ba_{p}+2\right)+h^{2}\left(a_{p},1,2\right)\right)\nonumber \\
&=\frac{1}{B}\left(\psi^{(1)}\left(Ba_{p}+2\right)+h^{2}\left(a_{p},1,2\right)\right)
\end{align}}%
 where $h$ is an auxiliary function defined as $h\left(a,i,j\right)=\psi\left(a+i\right)-\psi\left(Ba+j\right)$, and $\psi^{(k)}, h^{(k)}$ are the k-th derivative of the digamma function and the $h$ function, respectively. Finally, those moments are given by (\autoref{thm:pseudo-prob-moments}), 
{\begin{align}
\mu_{p} & =\frac{1}{B}\\
\sigma_{p}^{2} & =B\left(1-\frac{1}{B}\right)\left(\frac{1}{B^{3}}+\frac{\frac{4}{\pi}-1}{N}\left(\frac{1}{B^{2}}+\frac{1-\frac{1}{B}}{N}\left(\frac{1}{B}-\frac{1-\frac{2}{B}}{2N}\right)\right)\right)\end{align}}%
\end{proof}

\subsection{Hypothesis test}

To compare the modulation index of two signals of length $N$, we need to establish a null hypothesis that quantifies the chances of the metric (some discrepancy or distance between the modulation indices) being caused by chance. Therefore, we need to establish a distribution of MI values that are generated under a null case where there is no real coupling. In this specific aim, we assume this null model as a setting where two signals are independent white noise. Even though these two signals are not subject to any real phase-amplitude coupling, they have a non-zero MI that will vary with the number of bins and the length of compared signals.

Then, we propose to construct a statistic $\rho^*_{MI}$ based on the modulation index obtained from two white noise signals, $X$ and $Y$. This statistic has an approximated distribution described by \autoref{thm:modulation-index-distribution}. Therefore, we can calculate a p-value as the probability, under the null, of observing an MI that is at least as large as the metric that was observed. We can reject the null hypothesis if the p-value is higher than the significance level (typically, $\alpha=0.05$).

\section{Data analysis}

\subsection{Model validation}

The approximate distribution of the modulation index was determined on the basis of two main assumptions. First, the duration of each compared time series was assumed to be sufficiently long that the sum of Equation~\ref{eq:sim-H1} and Equation~\ref{eq:sim-H2} would be approximately equal to one. Second, it was postulated that the intermediate distributions precisely depict the statistical properties of the different stages while computing the modulation index. While the first assumption can be controlled in an experiment design phase by ensuring that the compared signals have more than 200 points (\autoref{sec:sum-of-weights-HT}), we needed to evaluate the validity of second conjecture. To pursue this goal, we conducted 100000 Monte Carlo simulations of modulation indices generated by white noise, using varying numbers of bins and sample lengths. The found empirical distributions were then compared with the proposed approximations.

\subsection{Applications in simulated data}

Furthermore, to illustrate the application of the hypothesis test in neural time series, we set up simulated scenarios with varying levels of cross-frequency coupling strength. Since the modulation index is sensitive to the number of bins, we also varied the number of bins in these scenarios to assess its effect. In particular, three scenarios were developed:

\textbf{Amplitude modulation (AM) scenario}. We simulated a low-frequency oscillation $x_{LF}(t)$ with a central frequency at 1Hz that modulated the amplitude of a high-frequency sinusoid (also known as a carrier) $x_{HF}(t)$ with a frequency at 20Hz. This type of coupling was widely used in analog communication to transmit electrical signals over long distances \citep{SigSystemsInference-Oppenheim-2016a}. The coupling strength in this scenario is the modulation power $A$, which measures the variation in amplitude in the carrier signal due to the low-frequency amplitude. Due to the frequencies of the modulating and carrier waves, we measure cross-coupling in the modulated signal between $0.1-5$Hz and $10-75$Hz. A similar simulated approach was used by Tsai et al. \citep{CmpEventrelatedModulation-Tsai-2022}.

\textbf{Spike-induced coupling scenario}. We simulated a single interictal epileptiform discharge as a sum of three scaled Gumbel probability distribution functions with durations of 60ms, 120ms, and 200ms which are representative of interictal epileptiform discharges \citep{EffectInterictalEpileptiform-Hu-2020}. To calculate the scale $\sigma$ of the Gumbel distribution, we define the duration $d$ as the interval where 95\% of the area of the function is contained, i.e., $\sigma=-d (\log(-\log(0.975)) -\log(-\log(0.025)))^{-1}$. Then the location $\mu$ was assumed to be in the middle of the duration. In this simulation, a spike was expected to appear every $\frac{2}{3}$ seconds. Each spike had a spectrum that ranges from 0Hz up to 20Hz, approximately, with a peak at 4Hz. We evaluated the modulation index between $0.1-8$Hz and $12-40$Hz. We chose the specific frequency ranges in this simulation such that portions of the spike spectrum will be shared across the low and high-frequency intervals that are compared. Therefore, we can assess the spurious contribution of spikes to the coupling assessment, with respect to our proposed significance level.

\textbf{High-frequency oscillation (HFO) coupling scenario}. HFOs are promising biomarkers to detect seizure onset zone in epileptic patients \citep{SimpleStatisticalMethod-Charupanit-2017} and it is often classified into two groups based on its frequency range: ripples (80–200 Hz) and fast ripples  (250-500 Hz) \citep{HighFreqOsc-Park-2019}. These types of oscillations are known to generate high levels of modulation indices \citep{AssessingEpileptogenicity-Bandarabadi-2019, PhaseAmplitudeCoupling-Li-2021}. We assessed the coupling magnitude due to a train of HFOs with respect to the 99\% critical value generated by our statistical model. Each HFO was modeled using a scaled Gaussian envelope with a duration of 200ms that modulates a 120Hz sinusoid. The duration was assumed to be the full-width-at-half-maximum, and therefore, the scale was calculated as $\sigma=\frac{d}{2.4}$. Likewise, it was simulated that the HFOs would appear at a comparable velocity to that of the modeled spikes. As expected, the HFO signals have a central frequency at 120Hz with a bandwidth of approximately  $114-126$Hz. Thus, we measure the cross-frequency coupling between $0.1-12$Hz and $90-147$Hz as the high-frequency and low-frequency components.

Each scenario was simulated for a length of 2 seconds, with a sampling frequency of 300Hz, and 9, 8, 36, and 60 bins were compared.

\section{Results}

\subsection{Monte Carlo simulations of white noise modulation index}

\begin{figure}
\hypertarget{fig:approximations}{%
\centering
\includegraphics{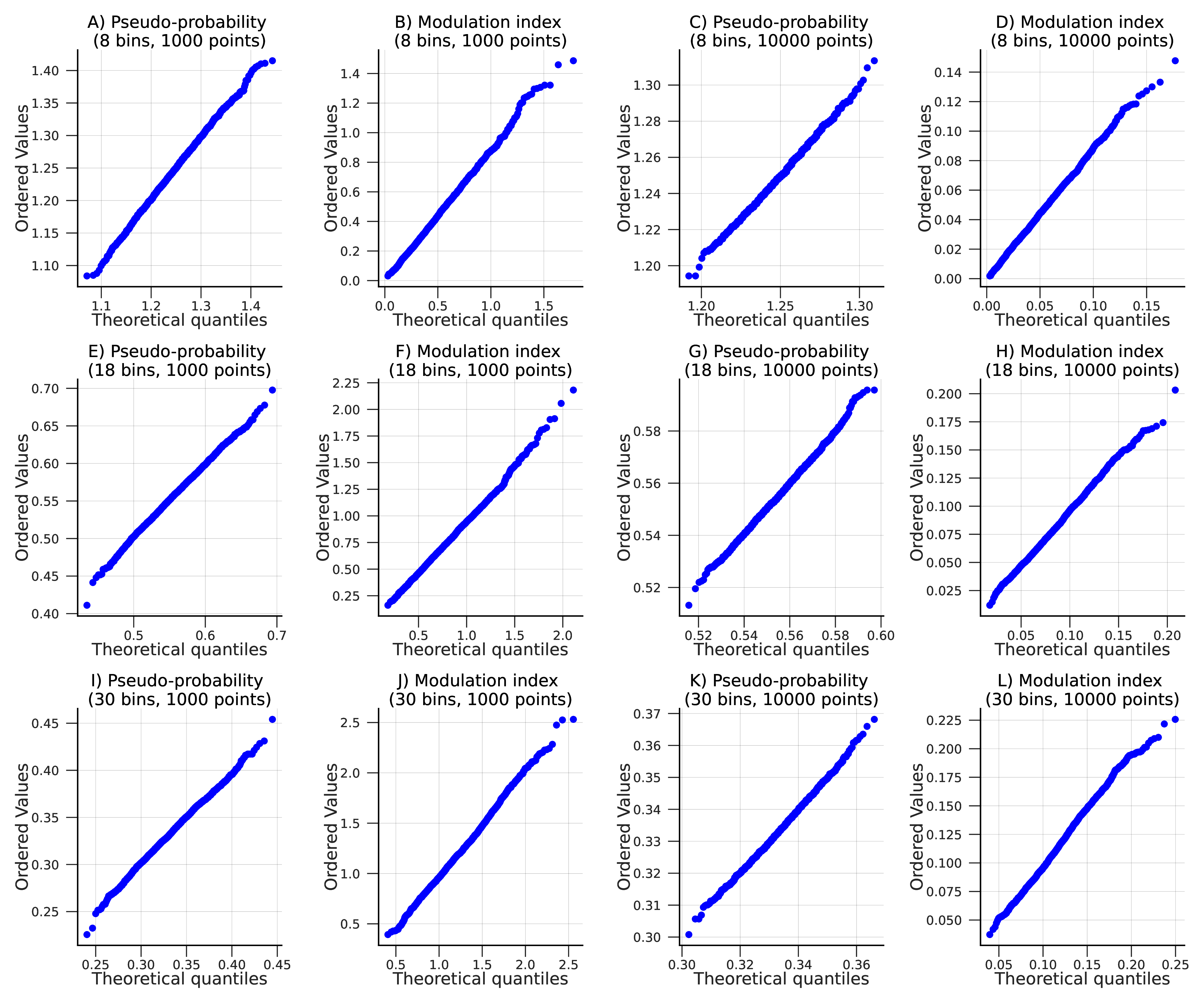}
\caption{Q-Q plots of the quantiles of the simulated MI values and the theoretical quantiles from our approximated model for the phase-amplitude histogram values ($P$) and the modulation indices $\rho_{MI}$. The comparison includes quantiles for 8, 18, and 30 bins with sample lengths of 1000 and 10000 points.}\label{fig:approximations}
}
\end{figure}

The first set of simulations shows that our proposed approximation reasonably fits the noise-generated modulation index. This goodness-of-fit can be assessed by the Q-Q plots shown in \autoref{fig:approximations}. Despite the evidence of some deviations in the extreme values, the theoretical distributions of the pseudo-probability and the modulation index are representatives of the simulated data, irrespective of the length of the sample and the number of bins.

\subsection{Modulation index in amplitude modulation}

Furthermore, the impact of the modulation in the different simulated scenarios provides an even deeper perspective on the utility of our statistical model. Recall that the theoretical lower bound given by the proposed statistical distribution provides a minimal expected modulation index that is not explained by noise and that this boundary can be affected by the number of bins (in the phase-amplitude histogram).% For each scenario, several values for coupling strength and the number of bins (in the phase-amplitude histogram) have been tested.

It is worth noting that in our context, the critical region pertains to the set of values at which our null hypothesis of no coupling is rejected, with a significance level of 99\%. The critical value is defined as the minimum value at which the null hypothesis cannot be rejected. Evidently, this critical value is equivalent to the value associated with p-value $p=0.01$, and it helps as a simple threshold to identify statistical significance.
In all three scenarios, as expected, the critical value increased as the number of bins increased, but it remained constant for all coupling scenarios: 0.00211 (8 bins), 0.00270 (18 bins), 0.00383 (36 bins), and 0.00527 (60 bins).

\begin{figure}
\hypertarget{fig:am}{%
\centering
\includegraphics{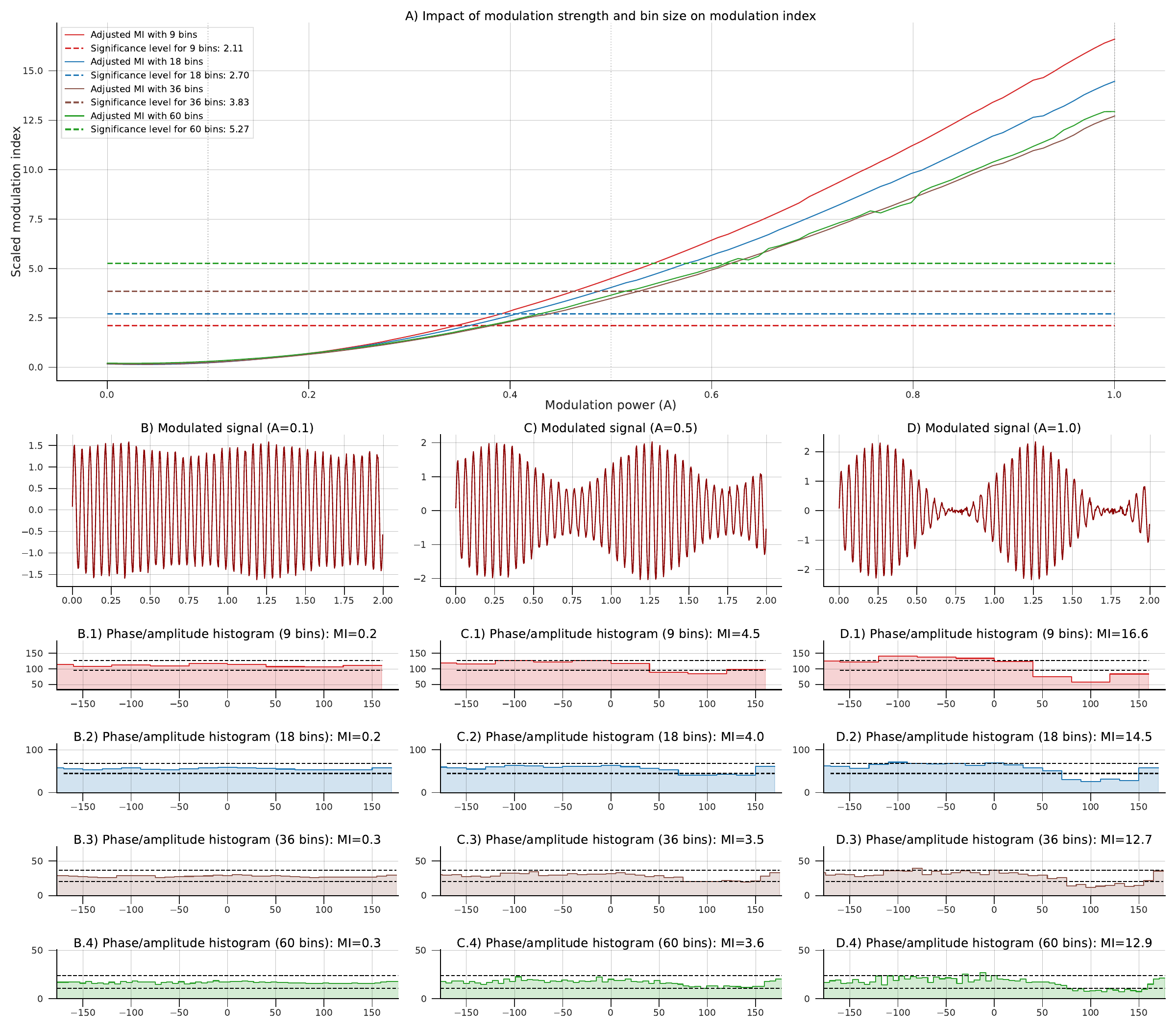}
\caption{Effect of coupling strength on modulation index A in an amplitude modulation scenario A) MI curves for 9, 18, 36, and 60 bins with strength ranging from 0 to 1. The dotted lines indicate the critical value at the 99\% confidence level, determined using our proposed null model. B) Minimal coupling (A=0.1) scenario with phase-amplitude histograms at various bin sizes. C) Scenario with moderate coupling (A=0.5). D) High coupling (A=1.0) scenario. The dotted lines in the histograms represent the 99\% confidence intervals determined using our null model.}\label{fig:am}
}
\end{figure}

In the case of amplitude modulation, we observe that the modulation power needed to be higher than 60\% to be outside of the critical region for all tested bins \autoref{fig:am}. However, if 18 bins are chosen to construct the phase-amplitude histogram, the threshold at which noise does not explain the obtained modulation index enables detection of coupling with modulation power higher than 40\%.

\subsection{Spurious coupling due to spikes}

\begin{figure}
\hypertarget{fig:spike}{%
\centering
\includegraphics{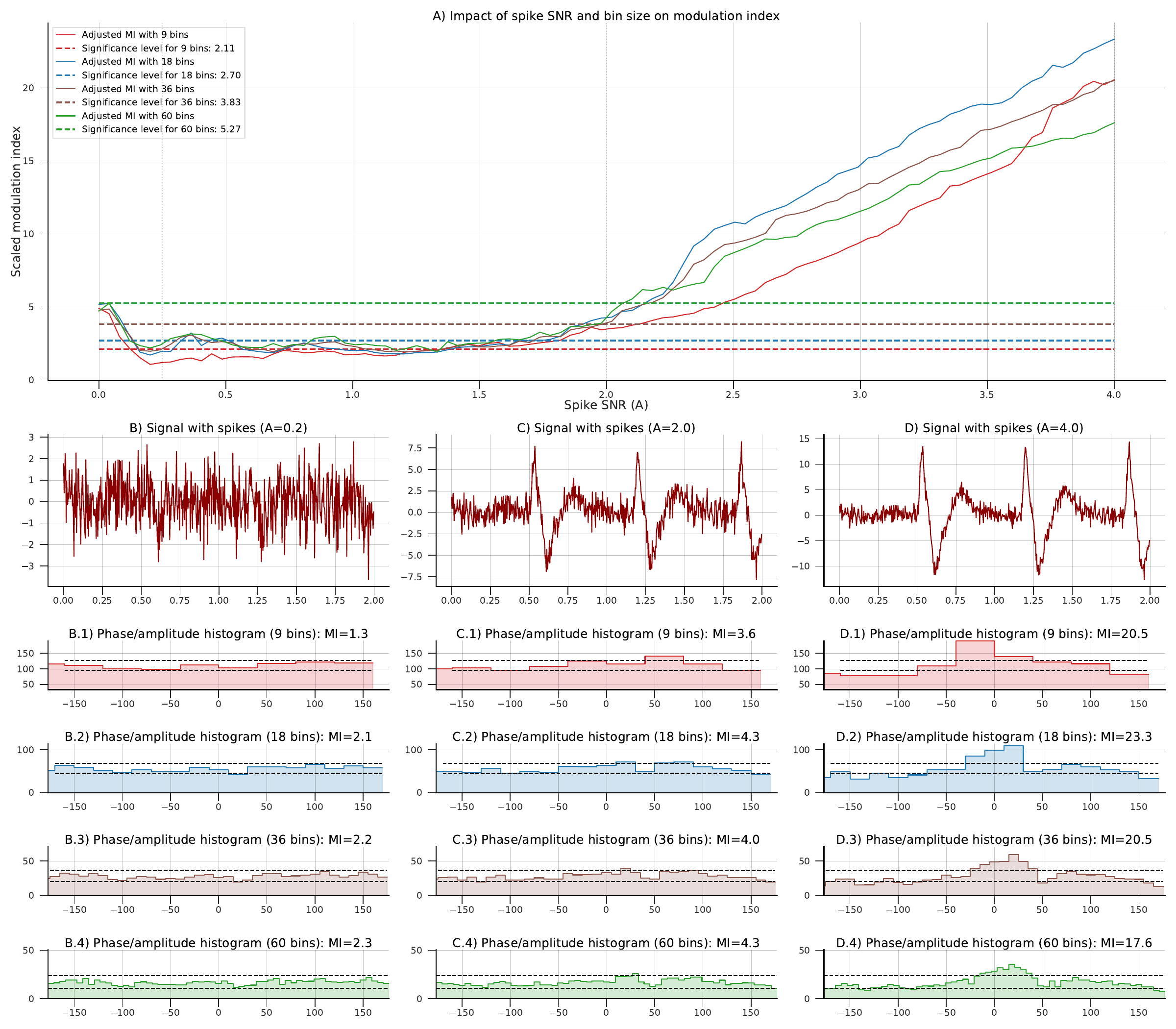}
\caption{Effect of signal-to-noise ratio (A) on modulation index in a train of spikes scenario A) MI curves are shown for 9, 18, 36, and 60 bins. The dotted lines indicate the critical value at the 99\% confidence level, determined using our proposed null model. B) Minimal coupling (A=0.2) scenario with phase-amplitude histograms at various bin sizes. C) Scenario with moderate coupling (A=2.0). D) High coupling (A=4.0) scenario. Each histogram has a 99\% confidence interval (dotted lines) of the values expected to see under noise. Note that only high events with high SNR deliver values outside this interval.}\label{fig:spike}
}
\end{figure}

The second simulated scenario showed the spurious coupling that a train of spikes can induce when the spectrum of interest coincides with the spectrum of the spikes. Compared with HFOs or the AM case, the spikes have a low-pass spectrum, and we expect a minimal effect in the PAC analysis if all frequencies of interest above approximately 30Hz. However, when we measure coupling betweem low-frequencies in the delta or theta band (0.1-8Hz) with medium-frequencies such as beta or gamma (12-40Hz), some noticeable effects are observed, with coupling effects higher than all tested critical thresholds when the SNR is higher than 2.5 \autoref{fig:spike}.

\subsection{Modulation index and HFO}

\begin{figure}
\hypertarget{fig:hfo}{%
\centering
\includegraphics{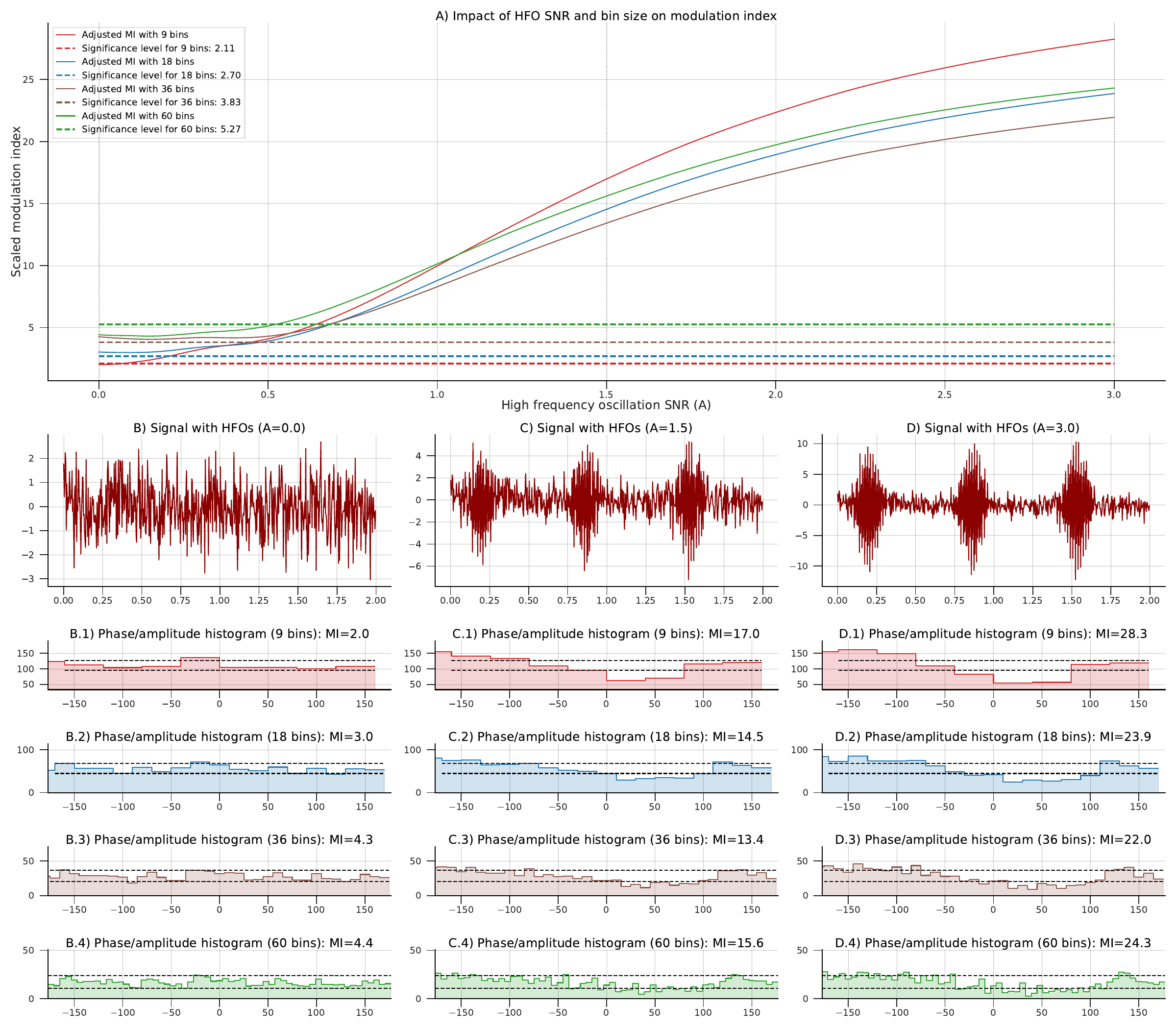}
\caption{Effect of signal-to-noise ratio (A) on modulation index for a train of HFOs A) MI curves are shown for 9, 18, 36, and 60 bins. The dotted lines indicate the critical value at the 99\% confidence level, determined using our proposed null model. B) No-coupling (A=0.0) scenario with phase-amplitude histograms at various bin sizes. C) Scenario with moderate coupling (A=1.5). D) High coupling (A=3.0) scenario. Each histogram is accompanied by a 99\% confidence interval (dotted lines).}\label{fig:hfo}
}
\end{figure}

The third analysis compared the coupling observed between the high-frequency components of a train of HFOs and low frequencies in the delta, theta, and alpha bands. The latter band was considered based on the 12Hz bandwidth of the simulated HFOs, which is presumed to be realistic. Under this consideration, it was observed that SNR higher than one was enough to surpass the critical thresholds for any tested bin \autoref{fig:hfo}. However, an SNR of around 0.25 was enough to be statistically significant when nine bins were used to calculate the modulation index. Compared with the other simulated scenarios, the MI in this simulation shows a higher divergence, bigger than 50e-3, in comparison with \textless10e-3, and \textless5e-3 in spikes and AM, respectively.

\subsection{Use of the critical value as cross-study comparisons}

As an illustration of the potential application of our statistical model for comparative analysis, we have gathered four studies that examined phase-amplitude coupling in animal models and reported the values they estimated: \citep{HippocampalThetaGamma-Jacobson-2013,LearningImprovesDecoding-Losacco-2020,PhaseamplitudeCoupledPersistent-Johnson-2017,VisualEvokedFeedforward-Aggarwal-2022}. These studies focused on interactions between theta (4-8Hz), low-gamma (30-50Hz), and high-gamma (65-100Hz) oscillations. We must emphasize that each study used different frequency ranges. However, our null statistical model only requires the number of sample points and the number of bins used in the MI analysis. These pieces of information have been compiled in \autoref{tab:commparison-cross-studies} along with the reported values and the critical value with a significance level of 0.01.

We should remark that \citep{PhaseamplitudeCoupledPersistent-Johnson-2017} reported that only MI values above 0.0001 would have been considered as relevant in their analysis. The exact procedure to generate this cut-off value was not fully specified. However, it coincided with the critical value of our null model at a significance level of 0.01 was, approximately, 0.0001094855.

\begin{table}[]
    \centering
    \begin{tabular}{>{\raggedright}m{0.12\columnwidth}>{\raggedright}m{0.1\columnwidth}>{\raggedright}m{0.1\columnwidth}>{\raggedright}m{0.4\columnwidth}>{\raggedright}m{0.16\columnwidth}}
     & Signal length & Number of bins & Reported MI & Critical value ($P=0.01$)\tabularnewline
    \hline 
    Jacobson et al., 2013 \citep{HippocampalThetaGamma-Jacobson-2013} & 7000 & 18 & Young rats$^A$: $(0.04,0.07)$
    
    Old rats$^A$: $(0.04,0.06)$
    
    Young rats$^B$: $(0.008,0.02)$
    
    Old rats$^B$: $(0.004,0.03)$ & $0.0002261$\tabularnewline
    \hline 
    Johnson et al., 2017 \citep{PhaseamplitudeCoupledPersistent-Johnson-2017} & 20000 & 36 & Max. MI value$^C$: $0.004$ & $0.0001095$\tabularnewline
    \hline 
    Losacco et al., 2019 \citep{LearningImprovesDecoding-Losacco-2020} & 24000 & 51 & Hippocampus$^A$: $(0.005, 0.030)$ & $0.0001105$\tabularnewline
    \hline 
    Aggarwal et al., 2022 \citep{VisualEvokedFeedforward-Aggarwal-2022} & 450 & 20 & Primary visual cortex$^C$: $0.00991$.
    
    Parietal cortex$^C$: $0.00155$.
    
    Max. over the stereotaxic grid$^C$: $0.019$ & $0.0046073$\tabularnewline
    \hline 
    \end{tabular}
    \caption{
    \label{tab:commparison-cross-studies}
    Comparison of reported modulation indices across studies. The experimental parameters, reported values, and the critical value associated with our null statistical model are provided with a significance level of 0.01. Reported MI values were calculated for different frequency intervals: theta-low gamma ($^A$), theta-high gamma ($^B$), theta-gamma ($^C$).
    }
    \label{tab:my_label}
\end{table}

\section{Discussion}

We proposed using a beta distribution to describe the statistical properties of MI values related to pure white noise. This approach allows us to determine the statistical significance of MI values via p-values or by setting thresholds below which MI values are deemed uninformative. It is reasonable to build a null model based on the MI values under white noise since it is thought that this kind of noise might reflect random fluctuations in brain activity. Even though MI critical values estimated by bootstrapping filtered or pink noise could have higher magnitudes \citep{EvaluationPhaseAmplitudeCoupling-Gohel-2016}, Gaussian-related critical values provide a minimal reference for assessing reliability in MI comparisons, and it has been applied by previous research \citep{EvaluationPhaseAmplitudeCoupling-Gohel-2016}, and studied as an explanatory process behind some types of cross-frequency interactions \citep{EnvelopeHuman-Hidalgo-2023}

We acknowledge that alternative approaches exist to construct a null distribution model from surrogate data. For instance, block resampling has been proposed to measure the distribution of coupling values that can be observed between time series that resemble the original ones but without any coupling effect. This resampling has been suggested for both amplitude and phase signals \citep{TimeresolvedPhaseamplitudeCoupling-Samiee-2017, UntanglingCrossfrequencyCoupling-Aru-2015, EvaluationPhaseAmplitudeCoupling-Gohel-2016}, only for phase time series \citep{AutomatedLateralizationTemporal-Gautham-2022, EmpiricalPhaseamplitude-Caiola-2019}, and only for the amplitude components \citep{HighGammaPower-Canolty-2006}. Block resampling essentially involves shuffling blocks of data instead of individual data points. However, it is important to note that this approach may introduce sudden ``phase jumps'' or change-point artifacts, as has been explored by Marshal et al. \citep{DetectionPhaseShift-Marshall-2014}. Such sudden changes have been known to induce spurious coupling \citep{SharpEdgeArtifacts-Kramer-2008}, which may lead to a bias in the null model and hence incorrect conclusions due to uncalibrated p-values.

The null model from which we developed our method is not affected by the aforementioned phase issues, and it requires minimal information from the signals to assess. Surprisingly, the variance of the data no longer plays a role in the estimation of the distribution after constructing the phase-amplitude histogram (\autoref{thm:pseudo-prob-moments}). In fact, the beta distribution we proposed in \autoref{thm:modulation-index-distribution} as an approximation for the null model only depends on the sample length and the number of bins used to construct such a histogram. We corroborated the accuracy of our approximated model with an extensive simulation of MI values generated by white noise. Our distribution offered the advantage of representing the stochastic information of MI indices with a minimal computational cost compared to shuffling the data or other types of simulations.

One of the limitations of MI is the lack of a direct interpretation of its values \citep{DirectModulationIndex-Scherer-2023}. A 99\% critical value from our statistical model could provide meaningful information to improve the interpretation of the results. We analyzed simulated scenarios that mimic realistic coupling effects while allowing for varying levels of coupling strength and described how the critical value could aid potential interpretations.

The results of our AM simulated scenario are consistent with the simulation and conclusion of Tsai et. who stated that MI remains stable over varying noise levels \citep{CmpEventrelatedModulation-Tsai-2022}. Notably, the index grows in a polynomial degree with respect to the strength of coupling. However, although the levels are continuously increasing, small modulation indices observed under weak modulation strength are susceptible to being masked by noise, as our statistical model showed. As expected, the critical value of the null hypothesis increased as the number of bins increased.

The spike simulation investigated the potential impact of spikes on coupling between the delta-theta and beta-low gamma frequency ranges, along with the capability of our model to set a reasonable level for rejection of MI values with low coupling effects. In general, sharp edges could increase MI estimations \citep{SharpEdgeArtifacts-Kramer-2008}. But, we must emphasize that although these situations commonly arise due to artifacts, they can also be caused by physiological responses. For instance, patients with pharmacoresistant epilepsy exhibit a high coupling level during NREM sleep in the frequency intervals used in our simulation \citep{HeterogeneousProfilesCoupled-Cox-2019}. Moreover, coupling in the delta band to higher frequencies could be linked to the sharp wave peaks of the ripples generated in the hippocampus as part of a control mechanism of slow neocortical oscillations \citep{RipplesMedialTemporal-Axmacher-2008} and can be observed consistently in intracranial electroencephalograms during NREM sleep \citep{RipplesMedialTemporal-Axmacher-2008}. High MI values can be observed if the frequencies of interest coincide with the spectrum bandwidth of spikes, and this magnitude can increase along with the spike's SNR with respect to the background. Our simulation showed that our MI's critical value was reasonably robust to reject spikes with a medium SNR magnitude. However, this implies that coupling generated with weak physiological-generated spikes can also be ignored.

%InterictalScalpFast-Bernardo-2020
%InteractionSlowWaves-Ellenrieder-2016
%DefiningRegionsInterest-Guirgis-2015
The last simulation involves the assessment of critical values to evaluate phase-coupling during pulses of high-frequency oscillations in the frequency intervals 0.1-4Hz and 90-147Hz. Previous studies remarked on the physiological relevance of MI associated with HFOs in similar intervals.
When comparing coupling between low-frequency oscillations (0.2-10Hz) and 50-100Hz HFOs, high MI values in the  temporal lobe have been suggested to be potential biomarkers of temporal lobe epilepsy in electrocorticograms, as MI is higher in mid-seizure epochs (five seconds before and after the middle point of a seizure interval)\citep{TemporalspatialPhaseamplitude-Zhang-2017}.%*
In rodent models, cross-frequency coupling in a similar frequency range of analysis was detected during sleep states due to the effects of antipsychotic medications (apomorphine and haloperidol) \citep{DeltamediatedCrossfrequencyCoupling-LopezAzcarate-2013}, but also in healthy rodents as a result of memory consolidation and neuromodulation processes \citep{InterregionalPhaseamplitudeCoupling-Atiwiwat-2023}. Our simulations were designed to represent realistic scenarios similar to the conditions in these studies but implemented such that the coupling strength could be controlled. Thus, the critical value of our statistical framework emphasized a minimal threshold that properly distinguished weak HFO-induced interactions.
A robust statistical assessment can improve the analysis of interactions involving HFOs, as the rate of these types of waves is associated with the seizure onset zones \citep{PhysRipplesAssociated-Bruder-2017,PhysRipplesAssociated-Bruder-2021}.

In addition, our null model allow us to compare results across studies performed in different experimental settings, with varying evaluated frequency intervals given the small number o parameters requires: number of points and number of bins used to estimate the phase-amplitude histogram \autoref{thm:pseudo-prob-moments}. As a analysis, we compile published results in \autoref{tab:commparison-cross-studies}. Several studies only provided the p-values \citep{ObjectiveInterictalElectrophysiology-Kuroda-2021,CouplingPhaseNeural-Wolpert-2021} of their statistical comparisons, their z-scores \citep{QuantificationPhaseAmplitudeCoupling-Hulsemann-2019}, or proportional MI values with unspecified factors \citep{TemporalSignatureSelf-Wolff-2019}, instead of the exact coupling values. The four studies in rodent models compiled in the aforementioned table provided a sufficient information about the observed scores and allows to test our null model.

In general, the reported MI values have magnitudes higher than our critical value by several orders of magnitude in \citep{HippocampalThetaGamma-Jacobson-2013,PhaseamplitudeCoupledPersistent-Johnson-2017,LearningImprovesDecoding-Losacco-2020}. The study in \citep{VisualEvokedFeedforward-Aggarwal-2022} was particularly interesting, because we could not reject the hypothesis that the theta-gamma MI values observed in the primary visual and parietal cortex were generated by random choice. However, we should emphasize that this do not imply the absence of phase-amplitude coupling, but that the coupling strength may have been influenced by noise or external artifacts given that higher MI values were observed in other regions in the same study.

From the experimental results, our statistical framework provides a meaningful advantage over existing approaches for evaluating the significance of modulation indices in phase-amplitude coupling. Unlike surrogate-based methods that require extensive computations \citep{PhaseAmplitudeCoupling-Tort-2010, AddressingPitfallsPhaseAmplitude-Jurkiewicz-2021, EvaluationPhaseAmplitudeCoupling-Gohel-2016}, our method provides a meaningful significance threshold (under a defined statistical significance level) with minimal computational requirements. This advantage is particularly noteworthy in cases where large replications of surrogate data are impractical due to the size of the dataset. However, despite the efficiency of our method, it is limited to the analysis of modulation indices caused by pure white noise. The significance levels of modulation indices generated by filtered or pink noise could be higher \citep{EvaluationPhaseAmplitudeCoupling-Gohel-2016}. As previously discussed, this bias also occurs with block-based surrogated methods. Despite these challenges, the proposed null model provides a robust mechanism for determining a minimal and reliable threshold, under a specific confidence level, that MI values must exceed to be considered for further analysis.

\section{Conclusions}

In this paper, we developed a statistical inferential framework for a null model approximation for the distribution of the modulation index (MI). Typically, a null model is obtained by creating bootstrapping from surrogated data. Our approach is to construct the null model by obtaining the distribution of observed MI values generated from white noise input. Here, we describe the closed-form probability distributions for all intermediate components involved in MI calculation, which enables us to characterize the null model of MI using a beta distribution that depends on a function of the number of sample points and bins used (\autoref{thm:modulation-index-distribution}).

We compare theoretical distributions with empirical distributions obtained from Monte Carlo simulations to validate our approach. Our theoretical approach captures the empirical quantile information accurately. Additionally, we create significance levels using these probability distributions to explore three different simulated scenarios that exhibit cross-frequency coupling phenomena found in neuroscience data: amplitude modulation, trains of spikes, and sequences of high-frequency oscillations. Our approach provides a reasonable minimum level of coupling associated with low levels of actual coupling for all three scenarios. However, we caution that those levels should not be considered an absolute reference due to inherent limitations associated with the MI formulation itself. Factors such as filtering effects or extreme values in the data could generate high spurious MI values, which could also be higher than the thresholds found by our statistical framework.

Our statistical framework for modulation indices offers a substantial advantage over alternative methods for evaluating the significance of MI values for phase-amplitude coupling events, as it can efficiently provide p-values or critical values under a defined statistical significance level. Our method is particularly beneficial in cases where obtaining large replications of surrogate data is impractical due to the size of the dataset or the number of events to compare. However, it is crucial to recognize that our approach is limited to analyzing modulation indices caused by pure white noise and that other types of noise (pink or filtered) may produce higher significance levels \citep{EvaluationPhaseAmplitudeCoupling-Gohel-2016}. Nevertheless, as the simulated events exhibited, our method can potentially reject extremely weak coupling phenomena that noisy interactions can explain.

To ensure reproducibility, we also include an open-source reference toolbox to evaluate the statistical significance of MI observations under our proposed model. The source code in different programming languages (Python, Javascript, R, and Matlab) is located in the Git repository \url{https://codeberg.org/mpinto/modulation-index-significance-assessment}. Thus, our approach could provide a reliable comparison framework suitable for meta-analyses and fast significance estimation in experimental settings.

\renewcommand{\thesection}{\Alph{section}}
 \setcounter{section}{0}

%
% APPENDICES
%
\begin{appendices}

\hypertarget{sec:analytic-signal-as-sum}{%
\section{Analytic signal}\label{sec:analytic-signal-as-sum}}

Recall that the analytic function is an operator over a signal $x(n)$ defined by %
\begin{equation}%
{\mathcal{A}\left\{ x\right\} (n)=\mathfrak{F}^{-1}\left\{ X\left(k\right)+\text{sgn}\left(k\right)X\left(k\right)\right\}}%
\end{equation}%
 where $\mathfrak{F}$ and $\mathfrak{F}^{-1}$ are the Fourier transform, and inverse Fourier transform, respectively, and $\text{sgn}\left(\cdot\right)$ is the sign function described by \begin{equation}{\text{sgn}\left(k\right)=\begin{cases}
1 & k\ge0\\
-1 & k<0
\end{cases}}%
\end{equation}%

Then, the analytic function is the sum of the signal itself and an imaginary component, $\hat{x}(n)$, also known as the Hilbert transform: %
\begin{equation}%
{\mathcal{A}\left\{ x\right\} (n)=x(n)+jh(n)\ast x(n)=x(n)+j\hat{x}(n)}%
\end{equation}%
 where $h(n)$ is a discrete Dirichlet kernel given by \begin{equation}{h(n)=\begin{cases}
0 & n\text{ even}\\
\frac{2}{\pi n} & n\text{ odd}
\end{cases}}%
\end{equation}%

Without loss of generality, let us assume that $N$ is even, such that the Hilbert transform can be expressed by %
\begin{equation}%
{\begin{aligned}\hat{x}(n) & =x(n)\ast h(n)\\
 & =\sum_{k=0}^{N-1}x\left(k\right)h\left(n-k\right)\\
 & =x\left(0\right)h\left(-n\right)+\ldots+x\left(n-2\right)h\left(-2\right)\\
 & \quad+x\left(n-1\right)h\left(-1\right)+x(n)h\left(0\right)+x\left(n+1\right)h\left(1\right)\\
 & \quad+x\left(n+2\right)h\left(2\right)+\ldots+x(n)h\left(N-n\right)
\end{aligned}}%
\end{equation}%
 Note that $x(n)h\left(0\right)=0$.

Then, we can reformulate the Hilbert transform as a weighted sum of future values subtracted by a weighted sum of previous values, %
\begin{equation}%
{\begin{aligned}\hat{x}(n) & =\sum_{\ell=1}^{n}x\left(n-\ell\right)h\left(-\ell\right)+\sum_{\ell=1}^{N-n}x\left(n+\ell\right)h\left(\ell\right)\\
 & =\sum_{\ell=1}^{N-n}x\left(n+\ell\right)h\left(\ell\right)-\sum_{\ell=1}^{n}x\left(n-\ell\right)h\left(\ell\right)
\end{aligned}}%
\end{equation}%

\hypertarget{sec:sum-of-weights-HT}{%
\section{Sum of weights in the Hilbert transform}\label{sec:sum-of-weights-HT}}

Let us explore some boundaries for the sum of the discrete Dirichlet kernel $h(n)$. For a signal with length $N$, the kernel sum from $0$ to $N$ is given by %
\begin{equation}%
{\begin{aligned}H_{1}(n)=\sum_{\ell=0}^{N}h(n) & =\frac{1}{2}\left(\psi\left(\frac{N}{2}+\frac{1}{2}\right)-\psi\left(\frac{1}{2}\right)\right)\frac{2}{\pi}\\
 & =\frac{1}{\pi}\left(\psi\left(\frac{N}{2}+\frac{1}{2}\right)-\psi\left(\frac{1}{2}\right)\right),\qquad N>0
\end{aligned}
\label{eq:sim-H1}}%
\end{equation}%
 where $\psi$ is the digamma function.

Similarly, the sum of the kernel's squared values kernel can be expressed as %
{\begin{align}\label{eq:sim-H2}H_{2}(n) & =\sum_{\ell=0}^{N}h^{2}(n)=\frac{1}{8}\left(\pi^{2}-2\psi^{(1)}\left(\frac{N}{2}+\frac{1}{2}\right)\right)\frac{4}{\pi^{2}}\\
 & =\frac{1}{2}\left(1-\frac{2}{\pi^{2}}\psi^{(1)}\left(\frac{N}{2}+\frac{1}{2}\right)\right),\qquad N>0
nonumber\end{align}
}%
 where $\psi^{(1)}$ is the first derivative of the digamma function.

Naturally, both expressions have a null value with $n=0$: $H_{1}(0)=H_{2}(0)=0$.

Therefore, numerically, we can obtain some upper bounds for the first derivative of the digamma function: %
\begin{equation}%
{\begin{aligned}\psi^{(1)}\left(k\right) & \le0.0675\qquad k\ge30\\
\psi^{(1)}\left(k\right) & \le0.02\qquad k\ge100\\
\psi^{(1)}\left(k\right) & \le0.01\qquad k\ge200
\end{aligned}}%
\end{equation}%

In turn, some boundaries can also be found for $H_{2}$: %
\begin{equation}%
{\begin{aligned}0.4932 & \le H_{2}(n)\le\frac{1}{2}\qquad k\ge30\\
0.4980 & \le H_{2}(n)\le\frac{1}{2}\qquad k\ge100\\
0.4990 & \le H_{2}(n)\le\frac{1}{2}\qquad k\ge200
\end{aligned}}%
\end{equation}%

\bibliographystyle{elsarticle-num}
%If additional references are needed, please add it to ExtraBib.bib (not in Library).
\bibliography{main}

\begin{thebibliography}{10}
\expandafter\ifx\csname url\endcsname\relax
  \def\url#1{\texttt{#1}}\fi
\expandafter\ifx\csname urlprefix\endcsname\relax\def\urlprefix{URL }\fi
\expandafter\ifx\csname href\endcsname\relax
  \def\href#1#2{#2} \def\path#1{#1}\fi

\bibitem{PhaseAmplitudeCouplingGeneral-Bergmann-2018}
T.~O. Bergmann, J.~Born, Phase-{{Amplitude Coupling}}: {{A General Mechanism}}
  for {{Memory Processing}} and {{Synaptic Plasticity}}?, Neuron 97~(1) (2018)
  10--13.
\newblock \href {https://doi.org/10.1016/j.neuron.2017.12.023}
  {\path{doi:10.1016/j.neuron.2017.12.023}}.

\bibitem{QuantificationPhaseAmplitudeCoupling-Hulsemann-2019}
M.~J. H{\"u}lsemann, E.~Naumann, B.~Rasch, Quantification of {{Phase-Amplitude
  Coupling}} in {{Neuronal Oscillations}}: {{Comparison}} of {{Phase-Locking
  Value}}, {{Mean Vector Length}}, {{Modulation Index}}, and
  {{Generalized-Linear-Modeling-Cross-Frequency-Coupling}}, Frontiers in
  Neuroscience 13 (2019) 573.
\newblock \href {https://doi.org/10.3389/fnins.2019.00573}
  {\path{doi:10.3389/fnins.2019.00573}}.

\bibitem{PhaseAmplitudeCoupling-Tort-2010}
A.~B.~L. Tort, R.~Komorowski, H.~Eichenbaum, N.~Kopell, Measuring
  {{Phase-Amplitude Coupling Between Neuronal Oscillations}} of {{Different
  Frequencies}}, Journal of Neurophysiology 104~(2) (2010) 1195--1210.
\newblock \href {https://doi.org/10.1152/jn.00106.2010}
  {\path{doi:10.1152/jn.00106.2010}}.

\bibitem{DirectModulationIndex-Scherer-2023}
M.~Scherer, T.~Wang, R.~Guggenberger, L.~Milosevic, A.~Gharabaghi, Direct
  modulation index: {{A}} measure of phase amplitude coupling for
  neurophysiology data, Human Brain Mapping 44~(5) (2023) 1862--1867.
\newblock \href {https://doi.org/10.1002/hbm.26190}
  {\path{doi:10.1002/hbm.26190}}.

\bibitem{ProperEstPhase-Dvorak-2014}
D.~Dvorak, A.~A. Fenton, Toward a proper estimation of phase\textendash
  amplitude coupling in neural oscillations, Journal of Neuroscience Methods
  225 (2014) 42--56.
\newblock \href {https://doi.org/10.1016/j.jneumeth.2014.01.002}
  {\path{doi:10.1016/j.jneumeth.2014.01.002}}.

\bibitem{TemporalspatialPhaseamplitude-Zhang-2017}
R.~Zhang, Y.~Ren, C.~Liu, N.~Xu, X.~Li, F.~Cong, T.~Ristaniemi, Y.~Wang,
  Temporal-spatial characteristics of phase-amplitude coupling in
  electrocorticogram for human temporal lobe epilepsy, Clinical
  Neurophysiology: Official Journal of the International Federation of Clinical
  Neurophysiology 128~(9) (2017) 1707--1718.
\newblock \href {https://doi.org/10.1016/j.clinph.2017.05.020}
  {\path{doi:10.1016/j.clinph.2017.05.020}}.

\bibitem{FRoleCrossfrequency-Canolty-2010}
R.~T. Canolty, R.~T. Knight, The functional role of cross-frequency coupling,
  Trends in Cognitive Sciences 14~(11) (2010) 506--515.
\newblock \href {https://doi.org/10.1016/j.tics.2010.09.001}
  {\path{doi:10.1016/j.tics.2010.09.001}}.

\bibitem{TransientPhaseamplitude-Martinez-Cancino-2019}
R.~{Mart{\'i}nez-Cancino}, J.~Heng, A.~Delorme, K.~{Kreutz-Delgado}, R.~C.
  Sotero, S.~Makeig, Measuring transient phase-amplitude coupling using local
  mutual information, NeuroImage 185 (2019) 361--378.
\newblock \href {https://doi.org/10.1016/j.neuroimage.2018.10.034}
  {\path{doi:10.1016/j.neuroimage.2018.10.034}}.

\bibitem{PhaseSynchrony-Lachaux-1999}
J.~P. Lachaux, E.~Rodriguez, J.~Martinerie, F.~J. Varela, Measuring phase
  synchrony in brain signals, Human Brain Mapping 8~(4) (1999) 194--208.
\newblock \href
  {https://doi.org/10.1002/(sici)1097-0193(1999)8:4<194::aid-hbm4>3.0.co;2-c}
  {\path{doi:10.1002/(sici)1097-0193(1999)8:4<194::aid-hbm4>3.0.co;2-c}}.

\bibitem{InfraslowOscModulate-Vanhatalo-2004}
S.~Vanhatalo, J.~M. Palva, M.~D. Holmes, J.~W. Miller, J.~Voipio, K.~Kaila,
  Infraslow oscillations modulate excitability and interictal epileptic
  activity in the human cortex during sleep, Proceedings of the National
  Academy of Sciences of the United States of America 101~(14) (2004)
  5053--5057.
\newblock \href {https://doi.org/10.1073/pnas.0305375101}
  {\path{doi:10.1073/pnas.0305375101}}.

\bibitem{PhaseAmplitudeReset-Mormann-2005}
F.~Mormann, J.~Fell, N.~Axmacher, B.~Weber, K.~Lehnertz, C.~E. Elger,
  G.~Fern{\'a}ndez, Phase/amplitude reset and theta-gamma interaction in the
  human medial temporal lobe during a continuous word recognition memory task,
  Hippocampus 15~(7) (2005) 890--900.
\newblock \href {https://doi.org/10.1002/hipo.20117}
  {\path{doi:10.1002/hipo.20117}}.

\bibitem{HighGammaPower-Canolty-2006}
R.~T. Canolty, E.~Edwards, S.~S. Dalal, M.~Soltani, S.~S. Nagarajan, H.~E.
  Kirsch, M.~S. Berger, N.~M. Barbaro, R.~T. Knight, High gamma power is
  phase-locked to theta oscillations in human neocortex, Science (New York,
  N.Y.) 313~(5793) (2006) 1626--1628.
\newblock \href {https://doi.org/10.1126/science.1128115}
  {\path{doi:10.1126/science.1128115}}.

\bibitem{AssessmentCrossfrequencyCoupling-Kramer-2013}
M.~Kramer, U.~Eden, Assessment of cross-frequency coupling with confidence
  using generalized linear models, Journal of Neuroscience Methods 220~(1)
  (2013) 64--74.
\newblock \href {https://doi.org/10.1016/j.jneumeth.2013.08.006}
  {\path{doi:10.1016/j.jneumeth.2013.08.006}}.

\bibitem{DynamicCrossfrequencyCouplings-Tort-2008}
A.~B.~L. Tort, M.~A. Kramer, C.~Thorn, D.~J. Gibson, Y.~Kubota, A.~M. Graybiel,
  N.~J. Kopell, Dynamic cross-frequency couplings of local field potential
  oscillations in rat striatum and hippocampus during performance of a
  {{T-maze}} task, Proceedings of the National Academy of Sciences of the
  United States of America 105~(51) (2008) 20517--20522.
\newblock \href {https://doi.org/10.1073/pnas.0810524105}
  {\path{doi:10.1073/pnas.0810524105}}.

\bibitem{AddressingPitfallsPhaseAmplitude-Jurkiewicz-2021}
G.~J. Jurkiewicz, M.~J. Hunt, J.~{\.Z}ygierewicz, Addressing {{Pitfalls}} in
  {{Phase-Amplitude Coupling Analysis}} with an {{Extended Modulation Index
  Toolbox}}, Neuroinformatics 19~(2) (2021) 319--345.
\newblock \href {https://doi.org/10.1007/s12021-020-09487-3}
  {\path{doi:10.1007/s12021-020-09487-3}}.

\bibitem{RoleDeltamodulatedHigh-Guirgis-2013}
M.~Guirgis, Y.~Chinvarun, P.~L. Carlen, B.~L. Bardakjian, The role of
  delta-modulated high frequency oscillations in seizure state classification,
  Annual International Conference of the IEEE Engineering in Medicine and
  Biology Society. IEEE Engineering in Medicine and Biology Society. Annual
  International Conference 2013 (2013) 6595--6598.
\newblock \href {https://doi.org/10.1109/EMBC.2013.6611067}
  {\path{doi:10.1109/EMBC.2013.6611067}}.

\bibitem{StatisticalFrameworkAssess-Nadalin-2019}
J.~K. Nadalin, L.-E. Martinet, E.~B. Blackwood, M.-C. Lo, A.~S. Widge, S.~S.
  Cash, U.~T. Eden, M.~A. Kramer, A statistical framework to assess
  cross-frequency coupling while accounting for confounding analysis effects,
  eLife 8 (2019) e44287.
\newblock \href {https://doi.org/10.7554/eLife.44287}
  {\path{doi:10.7554/eLife.44287}}.

\bibitem{EvaluationPhaseAmplitudeCoupling-Gohel-2016}
B.~Gohel, S.~Lim, M.-Y. Kim, K.-M. An, J.-E. Kim, H.~Kwon, K.~Kim, Evaluation
  of {{Phase-Amplitude Coupling}} in {{Resting State Magnetoencephalographic
  Signals}}: {{Effect}} of {{Surrogates}} and {{Evaluation Approach}},
  Frontiers in Computational Neuroscience 10 (2016) 120.
\newblock \href {https://doi.org/10.3389/fncom.2016.00120}
  {\path{doi:10.3389/fncom.2016.00120}}.

\bibitem{TimeresolvedPhaseamplitudeCoupling-Samiee-2017}
S.~Samiee, S.~Baillet, Time-resolved phase-amplitude coupling in neural
  oscillations, NeuroImage 159 (2017) 270--279.
\newblock \href {https://doi.org/10.1016/j.neuroimage.2017.07.051}
  {\path{doi:10.1016/j.neuroimage.2017.07.051}}.

\bibitem{UntanglingCrossfrequencyCoupling-Aru-2015}
J.~Aru, J.~Aru, V.~Priesemann, M.~Wibral, L.~Lana, G.~Pipa, W.~Singer,
  R.~Vicente, Untangling cross-frequency coupling in neuroscience, Current
  Opinion in Neurobiology 31 (2015) 51--61.
\newblock \href {https://doi.org/10.1016/j.conb.2014.08.002}
  {\path{doi:10.1016/j.conb.2014.08.002}}.

\bibitem{AutomatedLateralizationTemporal-Gautham-2022}
B.~K. Gautham, J.~Mukherjee, M.~Narayanan, R.~Kenchaiah, R.~C. Mundlamuri,
  A.~Asranna, V.~G. Lakshminarayanapuram, R.~D. Bharath, J.~Saini, C.~Nagaraj,
  S.~Mangalore, K.~Kulanthaivelu, N.~Sadashiva, A.~Mahadevan, J.~Rajan,
  K.~Kumar, A.~Arimappamagan, B.~R. Malla, S.~Sinha, Automated lateralization
  of temporal lobe epilepsy with cross frequency coupling using
  magnetoencephalography, Biomedical Signal Processing and Control 72 (2022)
  103294.
\newblock \href {https://doi.org/10.1016/j.bspc.2021.103294}
  {\path{doi:10.1016/j.bspc.2021.103294}}.

\bibitem{EmpiricalPhaseamplitude-Caiola-2019}
M.~Caiola, A.~Devergnas, M.~H. Holmes, T.~Wichmann, Empirical analysis of
  phase-amplitude coupling approaches, PLOS ONE 14~(7) (2019) e0219264.
\newblock \href {https://doi.org/10.1371/journal.pone.0219264}
  {\path{doi:10.1371/journal.pone.0219264}}.

\bibitem{HumanThalamusRegulates-Malekmohammadi-2015}
M.~Malekmohammadi, W.~J. Elias, N.~Pouratian, Human thalamus regulates cortical
  activity via spatially specific and structurally constrained phase-amplitude
  coupling, Cerebral Cortex (New York, N.Y.: 1991) 25~(6) (2015) 1618--1628.
\newblock \href {https://doi.org/10.1093/cercor/bht358}
  {\path{doi:10.1093/cercor/bht358}}.

\bibitem{RatiosNormalVariables-Marsaglia-1965}
G.~Marsaglia, Ratios of {{Normal Variables}} and {{Ratios}} of {{Sums}} of
  {{Uniform Variables}}, Journal of the American Statistical Association
  60~(309) (1965) 193--204.
\newblock \href {https://doi.org/10.1080/01621459.1965.10480783}
  {\path{doi:10.1080/01621459.1965.10480783}}.

\bibitem{SigSystemsInference-Oppenheim-2016a}
A.~V. Oppenheim, G.~C. Verghese, Signals, Systems \& Inference, {Pearson},
  {Boston}, 2016.

\bibitem{CmpEventrelatedModulation-Tsai-2022}
C.-C. Tsai, H.-H. Liu, Y.-L. Tseng, Comparison of event-related modulation
  index and traditional methods for evaluating phase-amplitude coupling using
  simulated brain signals, Biological Cybernetics 116~(5-6) (2022) 569--583.
\newblock \href {https://doi.org/10.1007/s00422-022-00944-7}
  {\path{doi:10.1007/s00422-022-00944-7}}.

\bibitem{EffectInterictalEpileptiform-Hu-2020}
D.~K. Hu, A.~Mower, D.~W. Shrey, B.~A. Lopour, Effect of interictal
  epileptiform discharges on {{EEG-based}} functional connectivity networks,
  Clinical Neurophysiology: Official Journal of the International Federation of
  Clinical Neurophysiology 131~(5) (2020) 1087--1098.
\newblock \href {https://doi.org/10.1016/j.clinph.2020.02.014}
  {\path{doi:10.1016/j.clinph.2020.02.014}}.

\bibitem{SimpleStatisticalMethod-Charupanit-2017}
K.~Charupanit, B.~A. Lopour, A {{Simple Statistical Method}} for the
  {{Automatic Detection}} of {{Ripples}} in {{Human Intracranial EEG}}, Brain
  Topography 30~(6) (2017) 724--738.
\newblock \href {https://doi.org/10.1007/s10548-017-0579-6}
  {\path{doi:10.1007/s10548-017-0579-6}}.

\bibitem{HighFreqOsc-Park-2019}
C.~J. Park, S.~B. Hong, High {{Frequency Oscillations}} in {{Epilepsy}}:
  {{Detection Methods}} and {{Considerations}} in {{Clinical Application}},
  Journal of Epilepsy Research 9~(1) (2019) 1--13.
\newblock \href {https://doi.org/10.14581/jer.19001}
  {\path{doi:10.14581/jer.19001}}.

\bibitem{AssessingEpileptogenicity-Bandarabadi-2019}
M.~Bandarabadi, H.~Gast, C.~Rummel, C.~Bassetti, A.~Adamantidis, K.~Schindler,
  F.~Zubler, Assessing {{Epileptogenicity Using Phase-Locked High Frequency
  Oscillations}}: {{A Systematic Comparison}} of {{Methods}}, Frontiers in
  Neurology 10 (2019) 1132.
\newblock \href {https://doi.org/10.3389/fneur.2019.01132}
  {\path{doi:10.3389/fneur.2019.01132}}.

\bibitem{PhaseAmplitudeCoupling-Li-2021}
Z.~Li, X.~Bai, R.~Hu, X.~Li, Measuring {{Phase-Amplitude Coupling Based}} on
  the {{Jensen-Shannon Divergence}} and {{Correlation Matrix}}, IEEE
  Transactions on Neural Systems and Rehabilitation Engineering 29 (2021)
  1375--1385.
\newblock \href {https://doi.org/10.1109/TNSRE.2021.3095510}
  {\path{doi:10.1109/TNSRE.2021.3095510}}.

\bibitem{HippocampalThetaGamma-Jacobson-2013}
T.~K. Jacobson, M.~D. Howe, B.~Schmidt, J.~R. Hinman, M.~A. Escab{\'i}, E.~J.
  Markus, Hippocampal theta, gamma, and theta-gamma coupling: Effects of aging,
  environmental change, and cholinergic activation, Journal of Neurophysiology
  109~(7) (2013) 1852--1865.
\newblock \href {https://doi.org/10.1152/jn.00409.2012}
  {\path{doi:10.1152/jn.00409.2012}}.

\bibitem{LearningImprovesDecoding-Losacco-2020}
J.~Losacco, D.~{Ramirez-Gordillo}, J.~Gilmer, D.~Restrepo, Learning improves
  decoding of odor identity with phase-referenced oscillations in the olfactory
  bulb, eLife 9 (2020) e52583.
\newblock \href {https://doi.org/10.7554/eLife.52583}
  {\path{doi:10.7554/eLife.52583}}.

\bibitem{PhaseamplitudeCoupledPersistent-Johnson-2017}
N.~W. Johnson, M.~{\"O}zkan, A.~P. Burgess, E.~J. Prokic, K.~A. Wafford, M.~J.
  O'Neill, S.~D. Greenhill, I.~M. Stanford, G.~L. Woodhall, Phase-amplitude
  coupled persistent theta and gamma oscillations in rat primary motor cortex
  in vitro, Neuropharmacology 119 (2017) 141--156.
\newblock \href {https://doi.org/10.1016/j.neuropharm.2017.04.009}
  {\path{doi:10.1016/j.neuropharm.2017.04.009}}.

\bibitem{VisualEvokedFeedforward-Aggarwal-2022}
A.~Aggarwal, C.~Brennan, J.~Luo, H.~Chung, D.~Contreras, M.~B. Kelz, A.~Proekt,
  Visual evoked feedforward\textendash feedback traveling waves organize neural
  activity across the cortical hierarchy in mice, Nature Communications 13~(1)
  (2022) 4754.
\newblock \href {https://doi.org/10.1038/s41467-022-32378-x}
  {\path{doi:10.1038/s41467-022-32378-x}}.

\bibitem{EnvelopeHuman-Hidalgo-2023}
V.~M. Hidalgo, J.~D{\'i}az, J.~Mpodozis, J.-C. Letelier, Envelope {{Analysis}}
  of the {{Human Alpha Rhythm Reveals EEG Gaussianity}}, IEEE Transactions on
  Biomedical Engineering 70~(4) (2023) 1242--1251.
\newblock \href {https://doi.org/10.1109/TBME.2022.3213840}
  {\path{doi:10.1109/TBME.2022.3213840}}.

\bibitem{DetectionPhaseShift-Marshall-2014}
W.~Marshall, P.~Marriott, Detection of {{Phase Shift Events}} (jan 2014).
\newblock \href {http://arxiv.org/abs/1401.3790} {\path{arXiv:1401.3790}}.

\bibitem{SharpEdgeArtifacts-Kramer-2008}
M.~A. Kramer, A.~B.~L. Tort, N.~J. Kopell, Sharp edge artifacts and spurious
  coupling in {{EEG}} frequency comodulation measures, Journal of Neuroscience
  Methods 170~(2) (2008) 352--357.
\newblock \href {https://doi.org/10.1016/j.jneumeth.2008.01.020}
  {\path{doi:10.1016/j.jneumeth.2008.01.020}}.

\bibitem{HeterogeneousProfilesCoupled-Cox-2019}
R.~Cox, T.~R{\"u}ber, B.~P. Staresina, J.~Fell, Heterogeneous profiles of
  coupled sleep oscillations in human hippocampus, NeuroImage 202 (2019)
  116178.
\newblock \href {https://doi.org/10.1016/j.neuroimage.2019.116178}
  {\path{doi:10.1016/j.neuroimage.2019.116178}}.

\bibitem{RipplesMedialTemporal-Axmacher-2008}
N.~Axmacher, C.~E. Elger, J.~Fell, Ripples in the medial temporal lobe are
  relevant for human memory consolidation, Brain: A Journal of Neurology
  131~(Pt 7) (2008) 1806--1817.
\newblock \href {https://doi.org/10.1093/brain/awn103}
  {\path{doi:10.1093/brain/awn103}}.

\bibitem{DeltamediatedCrossfrequencyCoupling-LopezAzcarate-2013}
J.~{L{\'o}pez-Azc{\'a}rate}, M.~J. Nicol{\'a}s, I.~Cordon, M.~Alegre,
  M.~Valencia, J.~Artieda, Delta-mediated cross-frequency coupling organizes
  oscillatory activity across the rat cortico-basal ganglia network, Frontiers
  in Neural Circuits 7 (2013) 155.
\newblock \href {https://doi.org/10.3389/fncir.2013.00155}
  {\path{doi:10.3389/fncir.2013.00155}}.

\bibitem{InterregionalPhaseamplitudeCoupling-Atiwiwat-2023}
D.~Atiwiwat, M.~Aquilino, O.~Devinsky, B.~L. Bardakjian, P.~L. Carlen,
  Interregional phase-amplitude coupling between theta rhythm in the nucleus
  tractus solitarius and high-frequency oscillations in the hippocampus during
  {{REM}} sleep in rats, Sleep 46~(4) (2023) zsad027.
\newblock \href {https://doi.org/10.1093/sleep/zsad027}
  {\path{doi:10.1093/sleep/zsad027}}.

\bibitem{PhysRipplesAssociated-Bruder-2017}
J.~C. Bruder, M.~D{\"u}mpelmann, D.~L. Piza, M.~Mader, A.~{Schulze-Bonhage},
  J.~{Jacobs-Le Van}, Physiological {{Ripples Associated}} with {{Sleep
  Spindles Differ}} in {{Waveform Morphology}} from {{Epileptic Ripples}},
  International Journal of Neural Systems 27~(07) (2017) 1750011.
\newblock \href {https://doi.org/10.1142/S0129065717500113}
  {\path{doi:10.1142/S0129065717500113}}.

\bibitem{PhysRipplesAssociated-Bruder-2021}
J.~C. Bruder, C.~Schmelzeisen, D.~{Lachner-Piza}, P.~Reinacher,
  A.~{Schulze-Bonhage}, J.~Jacobs, Physiological {{Ripples Associated With
  Sleep Spindles Can Be Identified}} in {{Patients With Refractory Epilepsy
  Beyond Mesio-Temporal Structures}}, Frontiers in Neurology 12 (2021) 612293.
\newblock \href {https://doi.org/10.3389/fneur.2021.612293}
  {\path{doi:10.3389/fneur.2021.612293}}.

\bibitem{ObjectiveInterictalElectrophysiology-Kuroda-2021}
N.~Kuroda, M.~Sonoda, M.~Miyakoshi, H.~Nariai, J.-W. Jeong, H.~Motoi, A.~F.
  Luat, S.~Sood, E.~Asano, Objective interictal electrophysiology biomarkers
  optimize prediction of epilepsy surgery outcome, Brain Communications 3~(2)
  (2021) fcab042.
\newblock \href {https://doi.org/10.1093/braincomms/fcab042}
  {\path{doi:10.1093/braincomms/fcab042}}.

\bibitem{CouplingPhaseNeural-Wolpert-2021}
N.~Wolpert, C.~{Tallon-Baudry}, Coupling between the phase of a neural
  oscillation or bodily rhythm with behavior: {{Evaluation}} of different
  statistical procedures, NeuroImage 236 (2021) 118050.
\newblock \href {https://doi.org/10.1016/j.neuroimage.2021.118050}
  {\path{doi:10.1016/j.neuroimage.2021.118050}}.

\bibitem{TemporalSignatureSelf-Wolff-2019}
A.~Wolff, D.~A. Di~Giovanni, J.~G{\'o}mez-Pilar, T.~Nakao, Z.~Huang,
  A.~Longtin, G.~Northoff, The temporal signature of self: {{Temporal}}
  measures of resting-state {{EEG}} predict self-consciousness, Human Brain
  Mapping 40~(3) (2019) 789--803.
\newblock \href {https://doi.org/10.1002/hbm.24412}
  {\path{doi:10.1002/hbm.24412}}.

\end{thebibliography}

\end{appendices}

\end{document}